\documentclass[letterpaper]{article} 
\usepackage{aaai24}  
\usepackage{times}  
\usepackage{helvet}  
\usepackage{courier}  
\usepackage[hyphens]{url}  
\usepackage{graphicx} 
\usepackage{url}            
\usepackage{hyperref}       
\usepackage{natbib}  
\usepackage{caption} 
\usepackage{amsmath,amssymb,amsfonts}
\usepackage{textcomp}
\usepackage{booktabs}

\usepackage{multirow}
\usepackage{subcaption}
\usepackage{tabularx}
\usepackage[linesnumbered,noend]{algorithm2e}
\usepackage{enumitem}
\usepackage{cleveref}
\usepackage{xspace}
\usepackage{adjustbox}
\usepackage{tikz}
\usetikzlibrary{tikzmark}
\usepackage{array}
\newcommand{\toolname}[0]{\texttt{\small RecXplainer}\xspace}

\newcommand{\baseline}{five\xspace}

\title{RecXplainer: Amortized Attribute-based \\ Personalized Explanations for Recommender Systems}

\author {
    Sahil Verma\textsuperscript{\rm 1} \thanks{Work done during an internship at Amazon},
    Chirag Shah\textsuperscript{\rm 1},
    John P. Dickerson\textsuperscript{\rm 2},
    Anurag Beniwal\textsuperscript{\rm 3},
    Narayanan Sadagopan\textsuperscript{\rm 3},
    Arjun Seshadri\textsuperscript{\rm 3},
}
\affiliations {
    \textsuperscript{\rm 1}University of Washington, Seattle\\
    \textsuperscript{\rm 2}University of Maryland\\
    \textsuperscript{\rm 3}Amazon\\
    vsahil@cs.washington.edu, chirags@uw.edu, john@cs.umd.edu, beanurag@amazon.com, sdgpn@amazon.com, sesarjun@amazon.com
}

\begin{document}
\maketitle

\thispagestyle{plain}
\pagestyle{plain}

\begin{abstract}
  Recommender systems influence many of our interactions in the digital world---impacting how we shop for clothes, sorting what we see when browsing YouTube or TikTok, and determining which restaurants and hotels we are shown when using hospitality platforms. Modern recommender systems are large, opaque models trained on a mixture of proprietary and open-source datasets. 
Naturally, issues of trust arise on both the developer and user side: is the system working correctly, and why did a user receive (or not receive) a particular recommendation? Providing an \emph{explanation} alongside a recommendation alleviates some of these concerns. 
The status quo for auxiliary recommender system feedback is either user-specific explanations (e.g., ``users who bought item B also bought item A'') or item-specific explanations (e.g., ``we are recommending item A because you watched/bought item B''). 
However, users bring personalized context into their search experience, valuing an item as a function of that item's attributes and their own personal preferences. 

In this work, we propose \toolname, a novel method for generating fine-grained explanations based on a user's preferences over the attributes of recommended items.
We evaluate \toolname on five real-world and large-scale recommendation datasets using five different kinds of recommender systems to demonstrate the efficacy of \toolname in capturing users' preferences over item attributes and using them to explain recommendations. We also compare \toolname to \baseline baselines and show \toolname's exceptional performance on ten metrics. 



\end{abstract}

\section{Introduction}
\label{sec:intro}
Recommender systems direct our attention, telling us what to look at---and what not to. They are deployed widely at platform companies such as Netflix, YouTube, Yelp, Amazon, and Shein. There are two main approaches to generating recommendations~\citep{Adomavicius2005Toward,recobroad1}: content-based~\citep{content1,content2} and collaborative filtering~\citep{collab-reco,colab1}. Content-based systems use item attributes and/or user preferences to recommend new items, while collaborative filtering (CF)\ uses the wisdom of the crowd, based on user ratings. Hybrid recommender systems aim to combine both approaches~\citep[see, e.g.,][]{Burke2004HybridRS}.
Modern methods are opaque in their performance and their servicing of an end-user's needs~\citep[see, e.g.,][]{survey-reco-xai}, which we focus on in this paper. 


\begin{figure}
    \centering
    \includegraphics[width=0.35\textwidth]{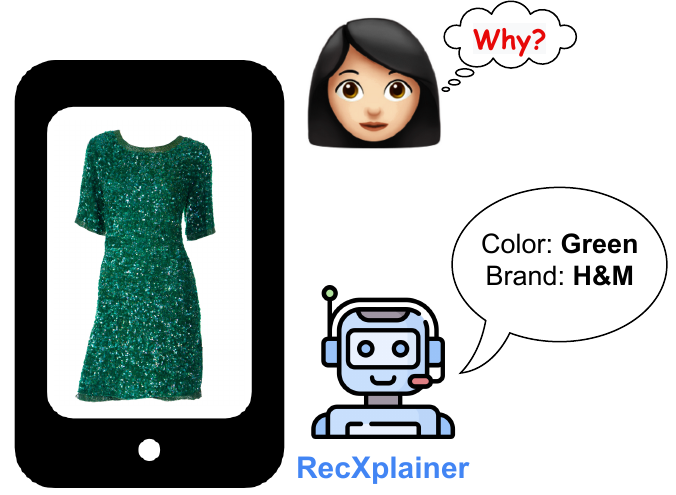}
    \caption{A user of a fashion website is recommended a cloth and she gets curious about the reason for recommending it? \toolname explains that it was recommended because she had shown interest in \textit{green} clothes and clothes from \textit{H\&M}. }
    \label{fig:recxplainer-in-action}
    \vspace{-0.7em}
\end{figure}

In recent years, CF-based methods have been widely chosen over content-based methods owing to 1) no requirement of manual labeling of item features, and 2) more serendipitous recommendations \citep{google-cf-serendipitious}. (Content-based recommender systems require each new item to be featurized, whose cost is usually high \citep{pandora-content-based}.)
However, CF-based methods have one major limitation: lack of interpretability. 
CF methods map users and items to an embedding space, which is learned from the user-item interaction matrix---and the proximity in this space is used to generate recommendations. Such an embedding space is difficult to interpret. A core challenge is understanding what a model learns about the user's preference over the items' attributes and explaining how it generates the new recommendations. 
\smallskip

Previous research has established that providing explanations for a recommendation enhances the transparency, scrutability, trustworthiness, and persuasiveness 
of the recommender systems \citep{Tintarev2007-thesis,xai-reco-motivation2,old-xai-reco-paper-sinha-2002}. 
This has spurred significant research in the broad field of ``explainability for CF recommender systems.'' 
Providing explanations when using CF-based methods is nontrivial due to the lack of interpretability of the embedding space. 
Most of the previous approaches provide explanations in the form of user-based or item-based explanations. User-based explanations explain a recommendation on the basis of `similar' users liking it. 
And item-based explanations explain a recommendation based on its similarity to other items that the user has previously liked. 
Item-based explanations are usually easier to grasp as the user knows about the items they have interacted with in the past. 
However, both of these explanation formats do not capture a user's preference over the attributes of an item, which is how users inherently think about a recommendation \citep{attribute-based-xai-reco2-radarchart-paper,attribute-based-xai-reco3-tagplanations-paper,attribute-based-xai-reco1,McAuley-and-Leskovec2013,McAuley2012LearningAttributes,rl-approach-xai-reco-MSFT}. 

This work proposes a novel approach, \toolname, that generates attribute-based explanations for CF-based recommender systems. 
A recommendation is explained in terms of a user's preference over the attributes of the item, (see \Cref{fig:recxplainer-in-action}), e.g., `We are recommending this movie because you like \emph{Action} movies.' or `We are recommending this t-shirt because you like \emph{Adidas} products and \emph{blue} clothes.' 
Such explanations are personalized to the user and hence help further enhance the persuasiveness and trustworthiness of the recommender systems. Note that the CF-based recommender system usually does not use these attributes to generate the recommendations. 

\toolname learns a model of the user's preference over the attributes of the items and uses this model to explain novel recommendations. The key advantages of \toolname are that it a) provides post-hoc explanations, b) is model-agnostic, and c) generates explanations in an amortized fashion. 
Post-hoc explanation techniques provide explanations after the recommender systems have been trained and therefore do not interfere with their architecture or training routine. 
This is crucial for real-world recommender systems whose architectures are complex, and whose training routines are highly optimized for accuracy. Moreover, such explainability techniques evade the accuracy-interpretability trade-off \citep{rl-approach-xai-reco-MSFT,Peake-and-Wang-data-mining2018}. 
Model-agnostic explanation techniques are desirable for their flexibility and generalizability as they can be used to explain a broad class of recommender systems \citep{molnar2022-xaibook}. 
Moreover, \toolname generates explanations in an amortized fashion, i.e., once \toolname learns the user's preference over attributes, it can generate explanations for novel recommendations by merely doing a model inference -- which is cheap and hence scalable to large recommender systems. 

We found that the important intersection of post-hoc attribute-based explanations for CF-based recommender systems has not garnered sufficient attention in explainability literature. 
We are aware of only two previous approaches that generate attribute-based explanations for CF-based recommender systems: LIME-RS \citep{limers-paper} and AMCF \citep{amcf-paper} (details in related work in \Cref{sec:related}). 
LIME-RS did not evaluate its attribute-based explanations, and the metrics used by AMCF were not generalizable to all CF-based recommender systems (see \Cref{sec:amcf-metrics-bad}). To this end, we propose a set of eight generalizable metrics to evaluate attribute-based explainability techniques for recommender systems (\Cref{sec:eval}) (one of our major contributions). We also consider the often overlooked popularity-based explanation methods: for attribute-based explanations, such an approach would choose the most popular attributes of a recommended item as an explanation. In this work, we implemented three popularity-based baselines for a holistic evaluation. 

In summary, our contributions are as follows:
\begin{enumerate}[leftmargin=4pt,topsep=0pt,itemsep=1pt]
    \item We propose \toolname, a novel, post-hoc, and model-agnostic technique to provide attribute-based explanations for recommender systems in an amortized fashion. To the best of our knowledge, \toolname \textbf{is the first technique} to have all these desirable properties. 
    
    \item We propose novel metrics to evaluate attribute-based explainability methods for CF-based recommender systems. 
    
    \item We perform extensive experimentation with five different classes of recommender systems trained on five large scale recommendation datasets. We demonstrate the efficacy of \toolname and its comparison with \baseline baselines using ten metrics. 
    
    \item We also explore the comparison of \toolname with the often overlooked popularity-based explanation methods, revealing surprising results. 
\end{enumerate}

\section{\texttt{RecXplainer}\xspace: Architecture \& Algorithm}
\label{sec:algo}

\begin{figure}[h]
    \centering
    \caption{\toolname's architecture. We train an auxiliary model that takes as input the user embeddings from the trained recommender system and the item attribute vector. }
    \includegraphics[width=\columnwidth]{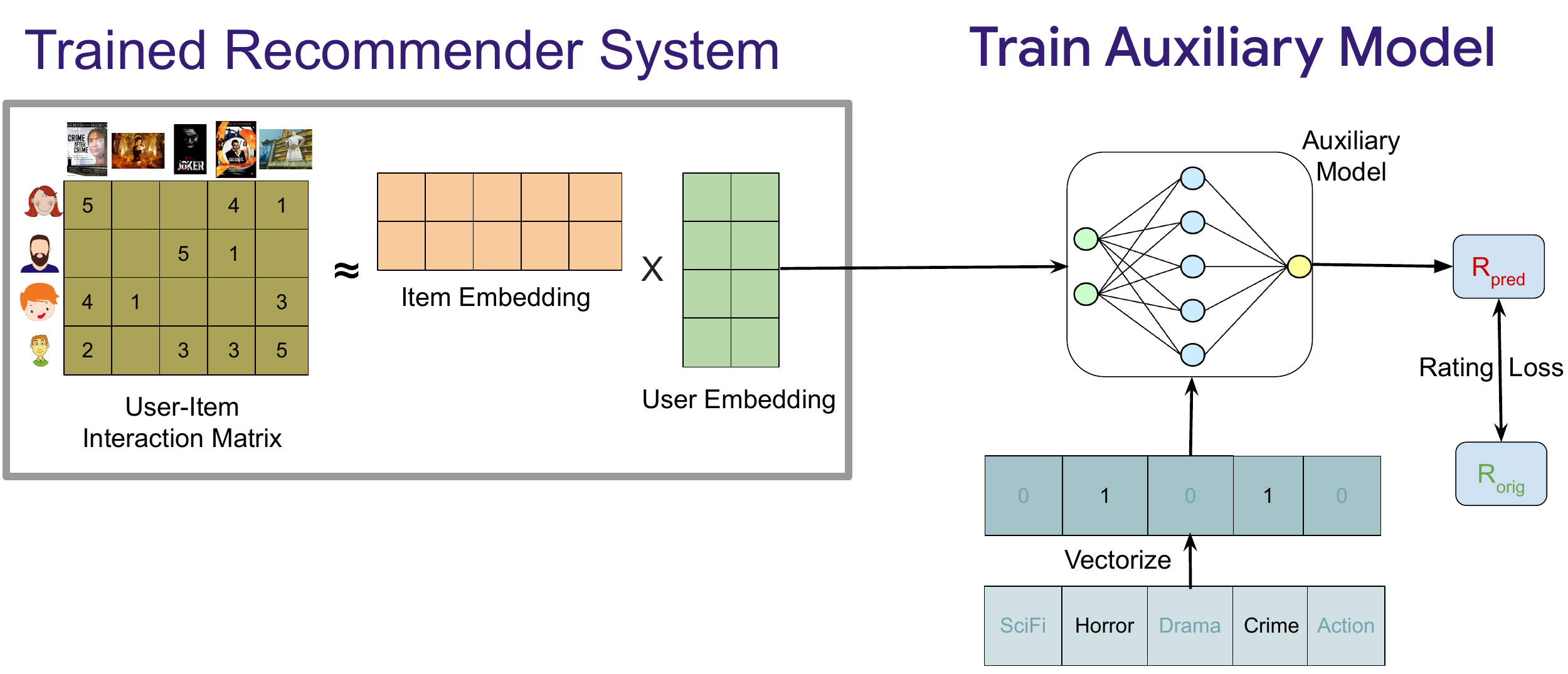}
    \label{fig:recxplainer-architecture}
    \vspace{-1.1em}
\end{figure}

\toolname provides explanations using the features of an item that a user prefers. In order to do so, we train an auxiliary model using the user embedding and the item's attribute vector. The user embedding is obtained from the trained recommender system's latent space and the item's attribute vector is constructed from the data. The auxiliary model is trained to reproduce the ratings given in the dataset (\emph{R\textsubscript{orig}} in the figure) using squared loss. \Cref{fig:recxplainer-architecture} shows \toolname's architecture. 

\paragraph{Item attribute vector:} The attribute vector is multi-hot -- it has 1s when the item has those attributes, otherwise, 0s. For example, consider a movie recommendation platform with metadata about the genres, and there are a total of 20 genres that all movies are categorized on. A particular movie will usually belong to 1-3 genres, and hence the attribute vector of a movie would consist of 1-3 1s in those indices (indicating the presence of those genres), and the rest of the indices will be 0s. Instead of binary values, the vector could also have continuous values. 

\paragraph{Auxiliary model architecture:} The auxiliary model can be instantiated with any model architecture. 
For experiments, we use three architectures: a linear layer, a neural network, and gradient boosted decision trees (GBDT). We trained the auxiliary models using Adam optimizer until convergence. Note that the auxiliary model is trained independently of the recommender system and it only requires the user's embedding from the recommender system. Therefore, \toolname is both post-hoc and model-agnostic. 

\subsection{\toolname: Generating Explanations}
\begin{figure}[h]
    \centering
    \caption{The difference in the scores produced by the auxiliary model for the cases when all features are present and a feature is removed is used to assign importance to the removed feature; thereby ranking the item features. }
    \includegraphics[width=\columnwidth]{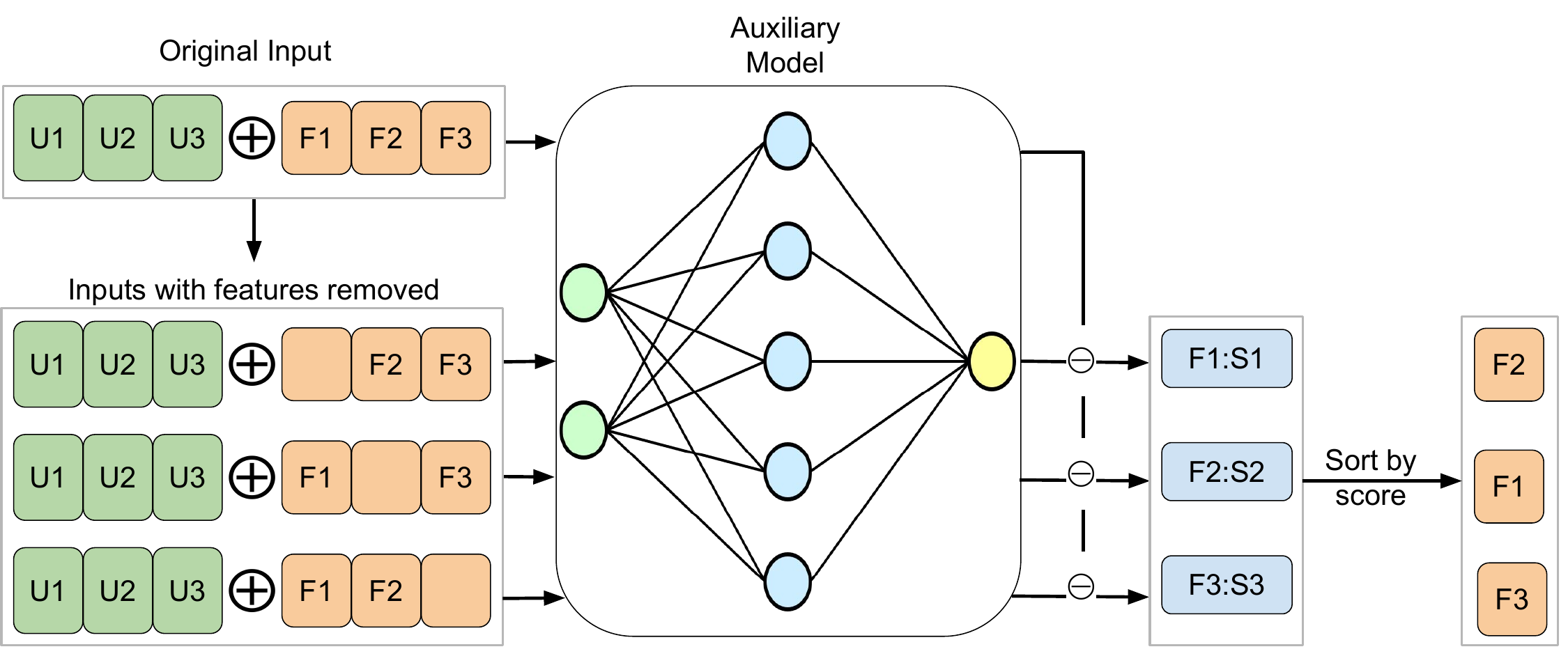}
    \label{fig:recxplainer-working}    
    \vspace{-1em}
\end{figure}

Once the auxiliary model is trained, we use it to generate explanations for new recommended items for the users (\Cref{fig:recxplainer-working}). 
Given that we need to explain an item recommended to a user (a user-item pair), \toolname produces a ranked order of that particular item's attributes in decreasing order of the user's preference over them. 
We term this as the user's \emph{specific preference} over the item's attributes. 
The importance of a particular attribute of an item is computed as the difference in the predicted ratings when the attribute is present and when it is absent from that item. The auxiliary model predicts both the ratings. 
For illustration, consider a movie recommendation platform. Alice is a user of the platform and gets a recommendation for a movie whose genres are \emph{Crime}, \emph{Romance}, and \emph{Documentary}. 
Alice is curious and requests an explanation: \toolname explains that she was recommended it because she likes \emph{Crime} movies. 

\vspace{4pt}
\begin{table}[h]
\centering
\caption{Illustration of how \toolname computes the specific preference of a user over an item's attributes}
\label{tab:specific-pref}
\resizebox{0.9\columnwidth}{!}{%
\begin{tabular}{ccc}
        \toprule
        Genre removed & Predicted rating & $\Delta$(Predicted rating) \\ 
        \midrule
        \textcolor{gray}{All genres present}  & \textcolor{blue}{4.2} \tikzmark{a} &  --  \\
        \emph{Crime}      & 3.2 \tikzmark{b} & \tikzmark{c} 1.0  \\
        \emph{Romance}    & 3.7 \tikzmark{d} & \tikzmark{e} 0.5  \\
        \emph{Documentary} & 3.9 \tikzmark{f} & \tikzmark{g} 0.3  \\ 
        \bottomrule
    \end{tabular}
    \begin{tikzpicture}[overlay, remember picture, shorten >=.5pt, transform canvas={yshift=.25\baselineskip}]
    \draw [->] ({pic cs:a}) [bend left] to ({pic cs:b});
    \draw [->] ([yshift=6pt,xshift=2pt]{pic cs:b}) -- ({pic cs:c});
    
    \draw [->] ({pic cs:a}) [bend left] to ({pic cs:d});
    \draw [->] ([yshift=6pt,xshift=2pt]{pic cs:d}) -- ({pic cs:e});
    
    \draw [->] ({pic cs:a}) [bend left] to ({pic cs:f});
    \draw [->] ([yshift=6pt,xshift=2pt]{pic cs:f}) -- ({pic cs:g});
  \end{tikzpicture}
}
\vspace{-0.6em}
\end{table}

Now let us look at how \toolname arrived at this explanation: \toolname requests the auxiliary model to predict the recommendation score for the same movie that the recommender system originally recommended. 
The auxiliary model predicts a rating of \textcolor{blue}{4.2} for this movie for Alice. It uses Alice's embedding and the movie's genres for this prediction. 
Next, \toolname approximates the importance of each individual attribute by zeroing them out one by one in the movie's genres vector. It then uses the auxiliary model to predict the rating of the movie had certain attributes not been present in the movie. 

\Cref{tab:specific-pref} shows this procedure. We zero out the genres present in the movie, one at a time, and see the change in the predicted rating (the original rating is the rating produced by the auxiliary model when all the genres of the movie are considered). The higher the drop in the predicted rating, the higher the importance of the removed genre, and hence the higher it's rank. 
In the above example, Alice's \emph{specific preference} over the genres is \emph{Crime}, \emph{Romance}, and \emph{Documentary}. \Cref{alg:spec-pref} provides the algorithm for predicting the \emph{specific preference} of a user for any item.

\SetKwProg{Fn}{Function}{}{}
\RestyleAlgo{algoruled}
\newcommand\mycommfont[1]{\fontsize{5.2}{5}\ttfamily\textcolor{blue}{#1}}
\SetCommentSty{mycommfont}

\newcommand{\xAlCapSty}[1]{\small\sffamily\bfseries\MakeUppercase{#1}}


\newcommand\mynlfont[1]{\scriptsize\sffamily{#1}}
\SetNlSty{mynlfont}{}{}
\SetAlFnt{\scriptsize\sf}

\SetSideCommentLeft
\vspace{-0.8em}
\begin{algorithm}[h]
 \DontPrintSemicolon 
 \SetKwInOut{KwIn}{Input}
 \SetKwInOut{KwOut}{Output}
 \SetKwData{importance}{SpecificPref}

 \KwIn{User Embedding $\mathcal{U}$, Item Features I, Trained Auxiliary Model $\mathcal{M}$}
 \KwOut{Preference of each attribute of the item \importance}
 
 \caption{ \label{alg:spec-pref} Predict the specific preferences of a user over the attributes of an item}
 
 \SetKwFunction{getscore}{get\_recommendation\_score}
 \SetKwFunction{getspecificpreference}{get\_specific\_preference}
 \SetKwFunction{sort}{sort}
 \SetKwData{score}{recommendation\_score}

    \Fn{ $ \mathit{\getscore(\mathcal{M}, \mathcal{U}, I')} $ }
    {
        $\mathit{\score \leftarrow \mathcal{M}(\mathcal{U} \oplus I')}$ \\
        \Return{$\mathit{\score}$}
    }
    
    \Fn{$\mathit{\getspecificpreference(\mathcal{M}, \mathcal{U}, I)}$}{
        $\mathit{\importance \leftarrow 0}$ \\
        \tcp{Iterate over all the attributes, and set the value of the attributes that are present to 0}
        \For{$\mathit{i \in I}$}{
            $\mathit{I' \leftarrow I}$ \\
            \If{$\mathit{i = 1}$}{
                $\mathit{I'[i] \leftarrow 0}$ \\
                $\mathit{\importance[i] \leftarrow \getscore(\mathcal{M}, \mathcal{U}, \mathit{I'})}$ \tcp{Get the recommendation score for this item when attribute $\mathit{i}$ is set to 0} 
            }
        }
        $\mathit{\importance \leftarrow \sort(\importance)}$ \tcp{sort in descending order}
        \Return{$\mathit{\importance}$}
    }
\end{algorithm}
\vspace{-0.7em}

\toolname's approach of attributing importance to genres falls in a well-studied domain of removal-based explanation methods \cite{Ians-paper-xai-by-removal-survey}, where the difference in the model's prediction before and after removal of a feature is approximation of its importance. 
There are several alternatives in the choice of the algorithm for removing features for attribution. 
SHAP \citep{shap-paper} removes all subsets of the features, gets the model's predictions for all those feature subsets and uses the Shapley value formula \citep{shapley:book1952} to assign importances. 
The alternative, which we choose for \toolname is one feature removal as it is much cheaper than SHAP to compute. 
However, \toolname is agnostic to the specific feature removal technique and works well with both one feature removal and SHAP. 
In \Cref{sec:removal-based}), we provide an empirical comparison of one feature removal and SHAP when used for feature attribution by \toolname. 
We find that the \toolname performs slightly better when using SHAP, however, SHAP is \textcolor{blue}{70} times more expensive than one feature removal technique, and hence not scalable to large datasets. 
Therefore, in the evaluation section, we use one feature removal for the results. 

\subsection{Global Attribute Importance}
Given the procedure to generate a user's \emph{specific preference} over the attributes of an item, we can capture a user's preference over all the attributes in the recommender system. 
We term this as a user's \emph{general preference}. 
\toolname computes a user's \emph{general preference} by averaging the importance of each attribute over all the items that a user liked in the past (Eq. \ref{eq:genpref}). 
Note that since most items have a small subset of all the attributes (a movie would usually belong to 1-3 genres from the total set of genres), the importance of the attributes not present in an item is set to be 0. 

\vspace{-0.5em}
\begin{equation}
    \small
    \begin{aligned}
        L & = \text{the set of items that user u liked} \\
        \text{GeneralPref}[u]  & \ = \ \frac{1}{|L|} \sum_{i \in L} \text{SpecificPref}[u][i]
    \end{aligned}
    \label{eq:genpref}
\end{equation}

\emph{Specific preference} and \emph{general preference} respectively are the analogs of \emph{local interpretability} and \emph{global interpretability} provided by popular methods like LIME \citep{ribeiro_why_2016} and SHAP \citep{shap-paper}. 

\section{Evaluation}
\label{sec:eval}

Our experiments characterize the quality of explanations by the number of liked items in the user history and the number of top-$k$ recommendations that can be explained by \toolname. 

\subsection{Datasets and Recommender Systems}
To demonstrate \toolname's scalability, we experiment with several popular recommender datasets of increasing sizes. Specifically, we use \underline{\textsc{Movielens-$100K$}} \citep{movielens-dataset}, \underline{\textsc{Hetrec}} \citep{grouplensHetRec2011}, \underline{\textsc{Anime}} \citep{kaggleAnimeRecommendations}, \underline{\textsc{Movielens-$20M$}} \citep{grouplensMovieLens20m}, and \underline{\textsc{YahooMusic}} \citep{yahoomusicdataset} datasets for our experiments. 
In order to explain the SOTA recommender system on these datasets, we train 5 popular recommender architectures on each of these datasets, and choose the best performing architecture as the recommender system to explain for each dataset. \Cref{tab:best-test-loss} shows the best performing architecture for each dataset (in bold). 
\begin{table}[h]
\caption{The test loss of each architecture for each dataset. }
\label{tab:best-test-loss}
\begin{center}
\resizebox{0.9\columnwidth}{!}{%
    \begin{tabular}{cccccc}
    \hline
        Dataset/Model      & MF & AutoE     & NCF        & FM    & Deep FM    \\ \hline
        Movielens-100K & \textbf{0.82}      & 1.1            & 0.87           & 0.85                    & 0.91                        \\ \hline
        Hetrec         & \textbf{0.57}      & 0.84           & 0.62           & 0.64                   & 0.70                         \\ \hline
        Anime          & 2.1                & 1.7            & \textbf{1.5}   & 1.8                    & 2.0                        \\ \hline
        Movielens-20M  & 0.95               & 0.83           & 0.63           & 0.82                   & \textbf{0.62}               \\ \hline
        YahooMusic     & \textbf{568}     & \textit{Timeout} & 613          & 658                    & 624                       \\ \hline
    \end{tabular}%
    }
\end{center}
\vspace{-1em}
\end{table}

Each of these datasets have features that \toolname uses to explain a recommendation. For example, there are of 18 genres in the \underline{\textsc{Movielens-$100K$}} dataset: \emph{\small Action, Adventure, Animation, Children's, Comedy, Crime, Documentary, Drama, Fantasy, Film-Noir, Horror, Musical, Mystery, Romance, Sci-Fi, Thriller, War, Western}. Each movie usually belongs to 1-3 genres. 
\Cref{sec:datasetdetails} provides the further details about the number of users, number of items, and the features for each dataset. 

For all the datasets, we used 70\% of the data for training and validating the recommender models and 30\% of the data as the test set for computing the metrics. 

\subsection{Baselines}
We compare \toolname to \baseline baseline approaches:
\begin{itemize}[leftmargin=0pt,itemsep=1pt,topsep=1pt]
    \item \emph{LIME-RS~\citep{limers-paper}: } This is the only previous post-hoc attribute-based explainability approach for CF-based recommender systems. It trains a regression model for every item it needs to explain; hence, this approach is not amortized. As we will see later in the experimental results, LIME-RS cannot scale to large datasets due to this problem. 
    
    \item \emph{AMCF~\citep{amcf-paper}: } This is another attribute-based explainability method for CF-based recommender systems; however this technique is not a post-hoc method. It trains a model while the recommender system is being trained and thus requires modification in its architecture. For a fair comparison, we adapted AMCF to be post-hoc: \emph{AMCF-PH} (details in \Cref{sec:amcf-post-hoc-adapt}). We froze the recommender system and trained AMCF's model till convergence. Note that unlike our approach, AMCF takes as input the item embedding and hence does not apply to recommender systems that do not construct item embeddings explicitly, for example, autoencoder and RBM-based recommender systems. 
    
    \item \emph{Global popularity: } This approach finds the most popular attributes across the entire dataset, i.e., the attributes that are the most popular among all the items that are liked across the entire dataset. For every recommended item, this technique would serve the same explanation. It is immediately evident that the explanations will not be personalized to the users and might be uninformative depending on the attributes 
    
    \item \emph{User-specific popularity: } This approach finds the most popular attributes across the items that a particular user liked in the dataset. These explanations are more personalized to the user than global popularity; however, they can still be uninformative if a common feature exists in all liked items. 
    
    \item \emph{Random: } This baseline is for control. It selects a random set of attributes for each item to be explained when computing both specific and general preferences. 
\end{itemize}

For \toolname, we train three auxiliary models in the experiments: a linear layer (RX-Linear), a 4-hidden layer neural network (RX-MLP), and gradient boosted trees (RX-GBDT). 
We repeated all experiments for five random splits of the dataset and report the mean and the standard deviation for all the metrics. 

\subsection{Metrics}
We measure performance using ten metrics. The metrics can be categorized into two broad categories: \textit{coverage} of recommended items and \textit{personalization} to the user. Previous work has mostly used coverage based metrics \cite{Adbollahi-and-Nasraoui-constrain,Peake-and-Wang-data-mining2018}. We argue about the insufficiency of coverage based metrics and propose a set of eight metrics that measure the personalization of the explanations to the users. These metrics can be used to evaluate any attribute-based explanation method for recommender systems. 
\begin{enumerate}[leftmargin=0pt,itemsep=1pt,topsep=1pt]
    \item \emph{Test set coverage: } This metric finds the percentage of the test set items whose attributes intersect with the top-$k$ general preferences of a user. For example, suppose a technique identifies the top-3 general preferences of a user Adam as \emph{Action}, \emph{Comedy}, and \emph{Sci-Fi}. If Adam likes a movie whose genres are \emph{Action} and \emph{Romance}, it is counted as covered by this technique as \emph{Action} had been identified as one of the preferred genres. Conversely, had the movie belonged to \emph{Crime} and \emph{Adventure} genres, it would not have been covered. We measure this metric only for the items that a user liked. We consider an item `liked' if a user has rated it $4$ or higher (when the dataset ratings are between 1 and 5) and $7$ or higher (when the dataset ratings are between 1 and 10). We only consider the top-$k$ general preferences for a user when measuring coverage, where k is about 0.2 times the number of attributes in the dataset. We report the mean coverage over all the users for this metric. 
    
    \item \emph{Top-$20$ recommendations coverage: } Similar to test set coverage, this metric finds the percentage of each user's top-$20$ recommended items whose attributes intersect with the top-$k$ general preferences of that user. We report the mean coverage over all the users for this metric. 
\end{enumerate}

\noindent  \textbf{Insufficiency of coverage based metrics:} Since a few attributes in most recommender datasets are very popular, i.e., they occur in almost all items: identifying such an attribute as a user's preference will provide a very high test set and recommendations coverage. However, those attributes can be inaccurate, unpersonalized, or uninformative as an explanation. For example, consider a movie recommendation dataset containing language as one of the attributes, where most movies are English movies. If an explanation technique provides `English' as an explanation for the recommended movie; that technique will get a 100\% test set and recommendations coverage while being uninformative and unpersonalized to the users. 
Therefore, we propose a new set of metrics to measure the personalization of the explanations. 

\begin{enumerate}[resume,leftmargin=0pt,itemsep=1pt,topsep=1pt]
    \item \emph{Personalization of the explanations: } 
    To measure the personalization of the explanations we require the ground-truth information about the users' preferences over the attributes. However, this information is usually missing from the datasets. 
    Therefore, we propose two measures that act as reasonable proxies for a user's attribute preference: 
    
    \begin{enumerate}[leftmargin=7pt,noitemsep,topsep=1pt]
        \item \emph{Conditional Probability of Liking given an attribute is present:} This computes the probability that a user likes an item, given that an attribute is present in it. This measure is the ratio of the number of times an attribute is present in the items a user likes (e.g., rated 4 or higher) and the number of times it is present in all the items the user rated. 
        
        \item \emph{Odds of Liking vs. Disliking given an attribute is present: } This computes the ratio of the number of times an attribute is present in an item the user likes (e.g., rated 4 or higher) and the number of times it is present in an item that the user dislikes (e.g., rated 2 or lower). 
        
    \end{enumerate}

    Both measures provide a proxy of users' general preference over attributes, i.e., a ranked order of the attributes. We use the training set for computing both these proxy measures. 
    Now, considering the two proxies as a ground truth of user preference over attributes, we report four personalization metrics for each of them: 
    \begin{enumerate}[leftmargin=0pt,itemsep=1pt,topsep=1pt]
        \item \emph{General preferences coverage:} This metric computes if there is any intersection between the top-$k$ attributes of a user's general preferences identified by a technique and the top-$k$ preferences identified by a proxy measure. We report the mean coverage over all the users for this metric for both the proxies. 
        
        \item \emph{General preferences ranking:} This metric measures the similarity between the ranking of the top-$k$ attributes of a user's general preferences identified by a technique and the top-$k$ preferences identified by a proxy measure. 
        We use rank-biased overlap (RBO) as a measure of similarity between the two ranked lists. RBO has several advantages over traditional rank similarity metrics like Kendall's Tau or Spearman's correlation (see \Cref{sec:rbo-just}). We compute this metric for all the items a user has liked and report the mean score over all the users. 
        
        \item \emph{Specific preferences coverage:} Similar to general preferences, we also measure if there is any intersection between the top-$k$ attributes of an item by the specific preferences of a user identified by a technique and the top-$k$ preferences identified by a proxy measure. For a user, we compute this metric for all the items they liked and report the mean coverage over all the users. 
        
        \item \emph{Specific preferences ranking:} This metric measures the similarity between the ranking of the top-$k$ attributes of an item by the specific preferences of a user identified by a technique and the top-$k$ preferences identified by a proxy measure. We use RBO for computing the similarity between the rankings. We compute this metric for all the items a user has liked and report the mean score over all the users. 
        
    \end{enumerate}
    \smallskip
    Given these these metrics reported per proxy measure, we get a total of eight metrics for computing the personalization of the explanations. 
\end{enumerate}

\begin{table*}[h]
	\caption{The test set coverage, top-k recommendations coverage, and the 8 explanation personalization metrics are reported here (averaged over all users) for the \underline{\textsc{Matrix Factorization}} model trained using \underline{\textsc{MovieLens-100k}} dataset. The mean and standard deviation value for each metric is computed over 5 random data splits. We compare 3 architecture choices for auxiliary modeling of RecXplainer. The best results are highlighted in bold. } 
	\begin{center}
	\resizebox{0.9\textwidth}{!}{%
		\begin{tabular}{lllllllll}
			\toprule
			Metrics & LIME-RS & AMCF-PH & GBL Popl. & User Popl. & Random & RX-Linear & RX-MLP & RX-GBDT \\
			\midrule
			Testset Coverage & 57.42 $\pm$ 0.47 & 49.62 $\pm$ 4.59 & \textbf{84.69 $\pm$ 0.18} & 77.6 $\pm$ 0.59 & 32.83 $\pm$ 0.57 & 60.74 $\pm$ 0.21 & 67.16 $\pm$ 0.59 & 63.01 $\pm$ 0.3 \\
			Recommendations Coverage & 67.08 $\pm$ 2.68 & 44.2 $\pm$ 2.58 & \textbf{79.12 $\pm$ 2.83} & 70.12 $\pm$ 2.05 & 26.72 $\pm$ 0.89 & 69.3 $\pm$ 2.4 & 65.23 $\pm$ 2.28 & 64.27 $\pm$ 2.09 \\
			\midrule
			CondProb Generalpref Coverage & 59.85 $\pm$ 0.56 & 54.57 $\pm$ 1.73 & 25.07 $\pm$ 0.58 & 41.51 $\pm$ 0.44 & 45.07 $\pm$ 0.91 & 64.77 $\pm$ 1.75 & 60.78 $\pm$ 1.59 & \textbf{71.24 $\pm$ 1.12} \\
			CondProb Generalpref Ranking & 13.99 $\pm$ 0.41 & 13.74 $\pm$ 0.39 & 5.04 $\pm$ 0.16 & 9.76 $\pm$ 0.23 & 11.64 $\pm$ 0.29 & 15.8 $\pm$ 0.34 & 15.68 $\pm$ 0.17 & \textbf{20.28 $\pm$ 0.3} \\
			CondProb Specificpref Coverage & 48.87 $\pm$ 0.25 & \textbf{49.97 $\pm$ 1.14} & 25.1 $\pm$ 0.58 & 41.44 $\pm$ 0.44 & 43.94 $\pm$ 0.5 & 49.78 $\pm$ 0.48 & 49.06 $\pm$ 0.22 & \textbf{51.23 $\pm$ 0.27} \\
			CondProb Specificpref Ranking & 51.24 $\pm$ 0.15 & 49.89 $\pm$ 0.32 & 6.29 $\pm$ 0.11 & 10.29 $\pm$ 0.25 & 11.71 $\pm$ 0.14 & 51.62 $\pm$ 0.36 & 52.59 $\pm$ 0.27 & \textbf{55.98 $\pm$ 0.22} \\
			Odds Generalpref Coverage & 69.31 $\pm$ 1.43 & 60.98 $\pm$ 1.46 & 47.42 $\pm$ 0.55 & 55.78 $\pm$ 0.84 & 44.39 $\pm$ 0.49 & 74.42 $\pm$ 1.05 & 72.47 $\pm$ 1.16 & \textbf{81.15 $\pm$ 0.55} \\
			Odds Generalpref Ranking & 23.25 $\pm$ 0.58 & 18.94 $\pm$ 0.8 & 17.72 $\pm$ 0.16 & 27.63 $\pm$ 0.4 & 11.25 $\pm$ 0.32 & 25.87 $\pm$ 0.33 & 29.3 $\pm$ 0.83 & \textbf{33.57 $\pm$ 0.39} \\
			Odds Specificpref Coverage & 54.96 $\pm$ 0.62 & 51.8 $\pm$ 0.93 & 47.47 $\pm$ 0.55 & 55.73 $\pm$ 0.84 & 44.08 $\pm$ 0.24 & 56.17 $\pm$ 0.5 & 55.56 $\pm$ 0.6 & \textbf{57.74 $\pm$ 0.46} \\
			Odds Specificpref Ranking & 50.67 $\pm$ 0.16 & 50.07 $\pm$ 0.44 & 19.04 $\pm$ 0.32 & 28.04 $\pm$ 0.4 & 11.68 $\pm$ 0.1 & 50.34 $\pm$ 0.12 & 52.54 $\pm$ 0.28 & \textbf{55.33 $\pm$ 0.12} \\
			\bottomrule
		\end{tabular}
	}
	\end{center}
	\label{tab:5-random-seeds-ml-100k-matrix_fact}
        \vspace{-1.3em}
\end{table*}

\subsection{Results and Discussion}

\Cref{tab:5-random-seeds-ml-100k-matrix_fact} reports the ten metrics for all the baselines and \toolname for the \emph{Matrix Factorization} based recommender system trained using \underline{\textsc{Movielens-$100K$}} dataset. We now analyze the results:
\begin{enumerate}[leftmargin=0pt,itemsep=1pt,topsep=1pt]
    \item \emph{LIME-RS: }  \toolname performs better than LIME-RS on all ten metrics. 
    
    \item \emph{AMCF-PH: } \toolname performs better than AMCF-PH on nine metrics and matches in one metric. 

    \begin{figure}[h!]
        \centering
        \includegraphics[width=0.7\columnwidth]{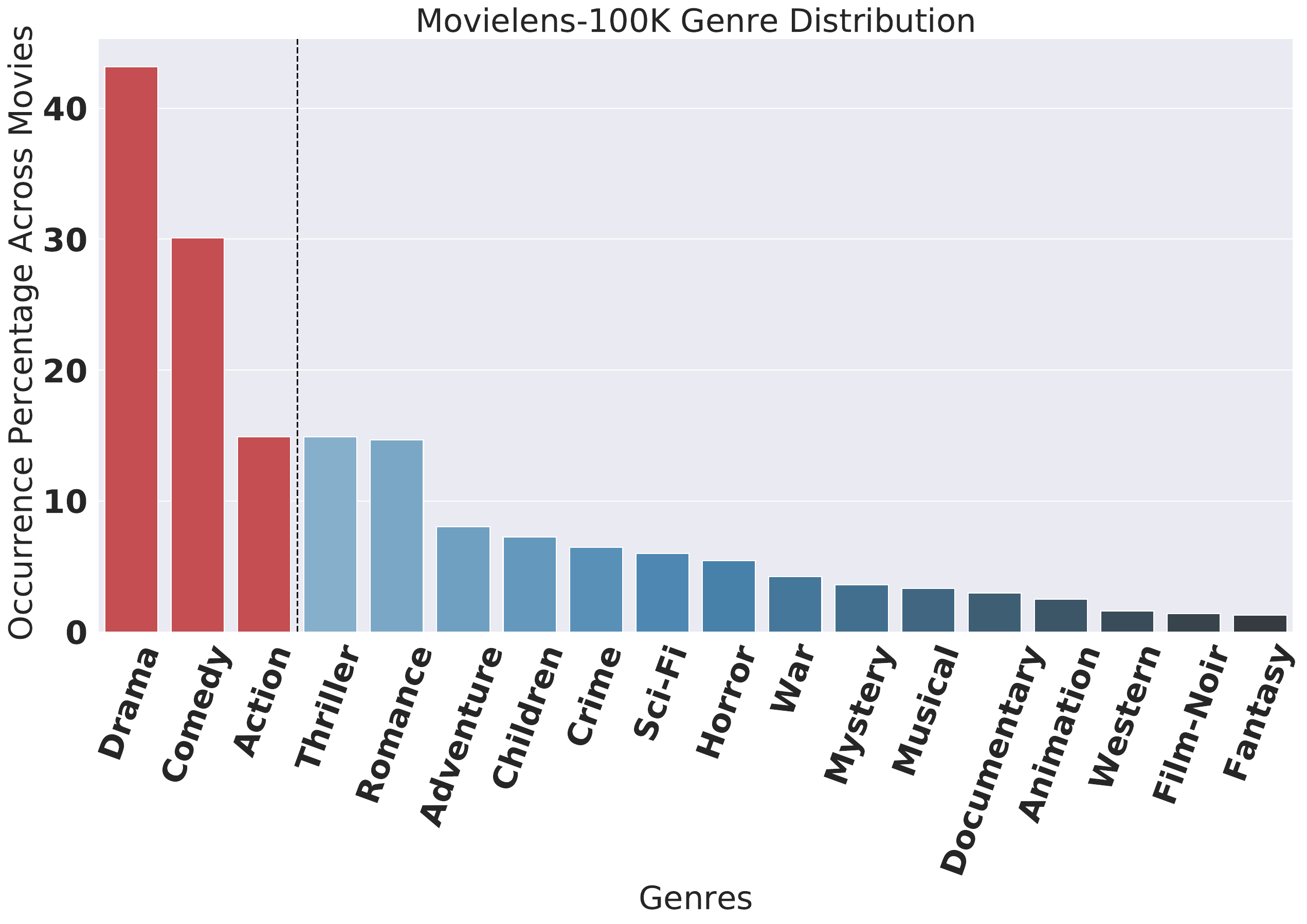}
        \label{fig:genre_dist_ml-100k}
    \end{figure}
    \vspace{-1.1em}

    \item \emph{Global popularity: } It achieves the highest test set and top-$k$ recommendations coverage. Upon further analysis, we found that the three most popular movie genres; \emph{Action}, \emph{Comedy}, and \emph{Drama} occur in over 88\% of the test set items (genre distribution plot in \Cref{fig:genre_dist_ml-100k}). Hence global popularity technique serves these three attributes as the explanation for all movies, thus providing unpersonalized and uninformative explanations. 
    This is corroborated by its poor performance on all personalization metrics; it performs worse than even the \emph{random} baseline for several metrics. We conclude that the explanations served by global popularity does not capture a user's preference over attributes and performs well for the two coverage-based metrics because of the skewed distribution of the attributes. 
    
    \item \emph{User-specific popularity: } This popularity approach achieves the second highest test set and top-$k$ recommendations coverage, and similar to \emph{global popularity} performs worse than \toolname on all personalization metrics, even worse than the \emph{random} baseline on several metrics. 
    
    \item \emph{Random: } We used this approach to serve as control and to ensure that no metric was trivial to perform well on. \toolname performs better than it on all ten metrics. 
    
    \item \emph{\toolname: } \toolname has the third highest test set and top-$k$ recommendations coverage while performing the best on all personalization metrics. We conclude that \toolname serves the most personalized explanations while still being able to explain a large proportion of test set items and top-$k$ recommendations. Moreover, \toolname can provide explanations in an amortized manner, providing explanations faster when deployed. 
\end{enumerate}

\begin{table*}[h]
	\caption{The test set coverage, top-k recommendations coverage, and the 8 explanation personalization metrics are reported here (averaged over all users) for the \underline{\textsc{Matrix Factorization}} model trained using \underline{\textsc{Hetrec}} dataset. The mean and standard deviation value for each metric is computed over 5 random data splits. We compare 3 architecture choices for auxiliary modeling of RecXplainer. The best results are highlighted in bold.} 
        \vspace{-2pt}
	\begin{center}
	\resizebox{0.9\textwidth}{!}{%
		\begin{tabular}{lllllllll}
			\toprule
			Metrics & LIME-RS & AMCF-PH & GBL Popl. & User Popl. & Random & RX-Linear & RX-MLP & RX-GBDT \\
			\midrule
			Testset Coverage & 68.52 $\pm$ 0.47 & 68.49 $\pm$ 0.78 & \textbf{94.4 $\pm$ 0.06} & 90.65 $\pm$ 0.09 & 45.51 $\pm$ 0.51 & 75.46 $\pm$ 0.34 & 80.98 $\pm$ 1.18 & 78.77 $\pm$ 0.18 \\
			Recommendations Coverage & 81.19 $\pm$ 0.9 & 55.49 $\pm$ 1.46 & \textbf{86.16 $\pm$ 1.75} & 82.99 $\pm$ 1.3 & 30.16 $\pm$ 1.33 & 81.4 $\pm$ 1.22 & 79.65 $\pm$ 2.59 & 78.98 $\pm$ 0.88 \\
			\midrule
			CondProb Generalpref Coverage & \textbf{87.37 $\pm$ 0.21} & 72.6 $\pm$ 2.05 & 37.46 $\pm$ 0.65 & 50.97 $\pm$ 0.44 & 62.06 $\pm$ 1.89 & 84.61 $\pm$ 0.42 & 78.63 $\pm$ 2.16 & 84.56 $\pm$ 0.45 \\
			CondProb Generalpref Ranking & 18.01 $\pm$ 0.19 & 14.57 $\pm$ 0.9 & 5.31 $\pm$ 0.21 & 8.91 $\pm$ 0.21 & 12.43 $\pm$ 0.76 & 17.72 $\pm$ 0.19 & 16.4 $\pm$ 0.62 & \textbf{19.55 $\pm$ 0.37} \\
			CondProb Specificpref Coverage & 60.7 $\pm$ 0.39 & 59.23 $\pm$ 0.45 & 37.42 $\pm$ 0.63 & 50.93 $\pm$ 0.43 & 62.26 $\pm$ 0.08 & \textbf{63.52 $\pm$ 0.45} & 59.87 $\pm$ 0.45 & 61.86 $\pm$ 0.47 \\
			CondProb Specificpref Ranking & 52.58 $\pm$ 0.05 & 50.25 $\pm$ 0.26 & 4.3 $\pm$ 0.2 & 6.59 $\pm$ 0.12 & 10.47 $\pm$ 0.04 & 54.26 $\pm$ 0.06 & 54.13 $\pm$ 0.47 & \textbf{57.46 $\pm$ 0.09} \\
			Odds Generalpref Coverage & \textbf{85.34 $\pm$ 0.4} & 75.75 $\pm$ 1.57 & 44.63 $\pm$ 0.47 & 53.48 $\pm$ 1.25 & 63.39 $\pm$ 0.91 & \textbf{85.15 $\pm$ 0.5} & 79.01 $\pm$ 2.28 & \textbf{84.79 $\pm$ 0.53} \\
			Odds Generalpref Ranking & 22.36 $\pm$ 0.23 & 19.1 $\pm$ 0.31 & 13.91 $\pm$ 0.23 & 20.52 $\pm$ 0.34 & 12.83 $\pm$ 0.32 & 23.32 $\pm$ 0.28 & 23.08 $\pm$ 0.48 & \textbf{26.0 $\pm$ 0.22} \\
			Odds Specificpref Coverage & 63.48 $\pm$ 0.31 & 62.92 $\pm$ 0.17 & 44.64 $\pm$ 0.47 & 53.51 $\pm$ 1.25 & 62.37 $\pm$ 0.09 & \textbf{65.78 $\pm$ 0.42} & 63.05 $\pm$ 0.38 & 64.86 $\pm$ 0.29 \\
			Odds Specificpref Ranking & 51.85 $\pm$ 0.09 & 50.8 $\pm$ 0.26 & 13.75 $\pm$ 0.32 & 19.54 $\pm$ 0.3 & 10.51 $\pm$ 0.05 & 53.49 $\pm$ 0.14 & 53.36 $\pm$ 0.35 & \textbf{56.35 $\pm$ 0.04} \\
			\bottomrule
		\end{tabular}
	}
	\end{center}
	\label{tab:5-random-seeds-hetrec-matrix_fact}
\end{table*}

\smallskip
\Cref{tab:5-random-seeds-hetrec-matrix_fact} reports the ten metrics for all the baselines and \toolname for the \emph{Matrix Factorization} based recommender system trained using \underline{\textsc{Hetrec}} dataset. We analyze the results:
\begin{enumerate}[leftmargin=0pt,itemsep=1pt,topsep=1pt]
    \item \emph{LIME-RS: }  \toolname performs better than LIME-RS for eight out of ten metrics and matches in one metric. 
    
    \item \emph{AMCF-PH: } \toolname performs better than AMCF-PH on all ten metrics. 
    
    \item \emph{Global popularity: } Similar to Movielens-$100K$, global popularity performs the best for the test set and top-$k$ recommendations coverage. This is due to the skewed distribution of genres (see \Cref{sec:attribute-dist-plots}). It performs worse than \toolname (even worse than the \emph{random} baseline) for all personalization metrics. 
    
    \item \emph{User-specific popularity: } Similar to Movielens-$100K$, user-specific popularity performs the second best for the test set and top-$k$ recommendations coverage, and worse than \toolname (even worse than \emph{random} baseline) for all personalization metrics. 
    
    \item \emph{Random: } \toolname performs better than the random baseline performs on all ten metrics. 
    
    \item \emph{\toolname: } \toolname performs the third best for test set coverage and the second best for top-$k$ recommendations coverage. It also performs the best for seven out of eight personalization metrics. This demonstrates that \toolname provides personalized explanations to the user while providing high coverage. 
\end{enumerate}

\begin{table*}[h]
	\caption{The test set coverage, top-k recommendations coverage, and the 8 explanation personalization metrics are reported here (averaged over all users) for the \underline{\textsc{Deep Factorization Machine}} model trained using \underline{\textsc{MovieLens-20M}} dataset. The mean and standard deviation value for each metric is computed over 5 random data splits. We compare 3 architecture choices for auxiliary modeling of RecXplainer. The best results are highlighted in bold.} 
        \vspace{-2pt}
	\begin{center}
	\resizebox{0.9\textwidth}{!}{%
		\begin{tabular}{lllllllll}
			\toprule
			Metrics & LIME-RS & AMCF-PH & GBL Popl. & User Popl. & Random & RX-Linear & RX-MLP & RX-GBDT \\
			\midrule
			Testset Coverage & \textit{Timeout} & 57.54 $\pm$ 1.33 & \textbf{93.78 $\pm$ 0.01} & 91.02 $\pm$ 0.01 & 47.28 $\pm$ 0.04 & 73.47 $\pm$ 0.33 & 83.51 $\pm$ 1.34 & 79.64 $\pm$ 0.14 \\
			Recommendations Coverage & \textit{Timeout} & 44.69 $\pm$ 2.55 & \textbf{82.74 $\pm$ 1.77} & 76.37 $\pm$ 0.71 & 35.66 $\pm$ 0.74 & 72.0 $\pm$ 0.82 & 75.55 $\pm$ 0.61 & 72.57 $\pm$ 0.78 \\
			\midrule
			CondProb Generalpref Coverage & \textit{Timeout} & 56.26 $\pm$ 2.71 & 33.39 $\pm$ 0.09 & 42.24 $\pm$ 0.12 & 64.86 $\pm$ 0.18 & \textbf{74.81 $\pm$ 0.54} & 63.26 $\pm$ 1.97 & 70.52 $\pm$ 0.18 \\
			CondProb Generalpref Ranking & \textit{Timeout} & 9.94 $\pm$ 1.02 & 4.33 $\pm$ 0.01 & 6.44 $\pm$ 0.01 & 13.15 $\pm$ 0.02 & \textbf{13.69 $\pm$ 0.21} & 11.08 $\pm$ 0.56 & \textbf{13.9 $\pm$ 0.08} \\
			CondProb Specificpref Coverage & \textit{Timeout} & 66.57 $\pm$ 0.89 & 33.38 $\pm$ 0.09 & 42.23 $\pm$ 0.12 & 64.78 $\pm$ 0.02 & \textbf{71.17 $\pm$ 0.73} & 64.49 $\pm$ 0.4 & 66.37 $\pm$ 0.11 \\
			CondProb Specificpref Ranking & \textit{Timeout} & 45.95 $\pm$ 0.43 & 3.44 $\pm$ 0.02 & 4.85 $\pm$ 0.01 & 11.11 $\pm$ 0.01 & 53.07 $\pm$ 0.11 & 51.31 $\pm$ 0.25 & \textbf{54.23 $\pm$ 0.05} \\
			Odds Generalpref Coverage & \textit{Timeout} & 62.48 $\pm$ 1.34 & 43.38 $\pm$ 0.05 & 50.98 $\pm$ 0.04 & 64.83 $\pm$ 0.17 & \textbf{83.88 $\pm$ 0.34} & 73.05 $\pm$ 2.06 & 79.68 $\pm$ 0.12 \\
			Odds Generalpref Ranking & \textit{Timeout} & 13.12 $\pm$ 0.5 & 12.2 $\pm$ 0.02 & 17.03 $\pm$ 0.02 & 13.15 $\pm$ 0.06 & 21.45 $\pm$ 0.11 & 20.3 $\pm$ 0.53 & \textbf{23.29 $\pm$ 0.03} \\
			Odds Specificpref Coverage & \textit{Timeout} & 63.68 $\pm$ 0.36 & 43.39 $\pm$ 0.05 & 50.97 $\pm$ 0.04 & 64.79 $\pm$ 0.01 & \textbf{68.09 $\pm$ 0.15} & 64.65 $\pm$ 0.22 & 66.17 $\pm$ 0.06 \\
			Odds Specificpref Ranking & \textit{Timeout} & 46.18 $\pm$ 0.49 & 11.99 $\pm$ 0.02 & 16.04 $\pm$ 0.02 & 11.11 $\pm$ 0.01 & 51.94 $\pm$ 0.05 & 51.03 $\pm$ 0.18 & \textbf{53.61 $\pm$ 0.01} \\
			\bottomrule
		\end{tabular}
	}
	\end{center}
	\label{tab:5-random-seeds-ml-20m-dfm}
        \vspace{-1.3em}
\end{table*}

\smallskip
\Cref{tab:5-random-seeds-ml-20m-dfm} reports the ten metrics for all the baselines and \toolname for the \emph{Deep Factorization Machine} based recommender system trained using \underline{\textsc{Movielens-$20M$}} dataset. We analyze the results:
\begin{enumerate}[leftmargin=0pt,itemsep=1pt,topsep=1pt]
    \item \emph{LIME-RS: } Recall that LIME-RS trains a separate regression model for each item to be explained. This is a very expensive procedure to generate explanations. For a large dataset like Movielens-20M that has 138,000 users, LIME-RS did not scale. LIME-RS took more than 1 hour to compute the general preferences for just 1000 randomly sampled users -- thereby, the estimated time for computing both the specific and general preferences for all users is \textit{over 11 days}. 
    
    \item \emph{AMCF-PH: } \toolname performs better than AMCF-PH on all ten metrics.  
    
    \item \emph{Global popularity: } Similar to the two previous datasets, global popularity performs the best for the test set and recommendations coverage, while performing worse than \toolname and even the \emph{random} baseline on all personalization metrics. 
    
    \item \emph{User-specific popularity: } Similar to the two previous datasets, user-specific popularity performs the second best for the test set and recommendations coverage, while performing worse than \toolname and even the \emph{random} baseline on all personalization metrics. 
    
    \item \emph{Random: } \toolname performs better than the random baseline performs on all ten metrics. 
    
    \item \emph{\toolname: } \toolname performs the third best on the two coverage metrics and performs the best on all eight personalization metrics, thereby demonstrating its ability to provide personalized explanations with high coverage. 
\end{enumerate}

\smallskip
Due to space constraints, \Cref{tab:5-random-seeds-anime-ncf} and \Cref{tab:5-random-seeds-yahoomusic-matrix_fact} in \Cref{sec:remainingdatasets} reports the metrics for \underline{\textsc{Anime}} and \underline{\textsc{YahooMusic}} datasets. The conclusion from those results is similar to the one from three recommender systems above. 

Overall, we conclude that \toolname performs better than all baselines for most metrics. When it is not the best, it is usually within a couple percentage points of the best performing technique. 
LIME-RS usually does not perform well on any metric and is unable to even scale to 3 out of 5 datasets owing to the expensive cost of training a separate regression model for each item. \toolname completely avoids this problem due to amortization, it is very cheap to produce \toolname explanations during deployment.  
AMCF-PH also does not perform well on most metrics for all datasets.  
Global and user-specific popularity perform well on the coverage-based metrics; however, that is due to the skewed distribution of the attributes, and their explanations are usually uninformative and unpersonalized. 

\subsection{Ablation Experiments}
We perform a series of ablations to demonstrate different aspects of \toolname:
\begin{enumerate}[leftmargin=4pt,itemsep=1pt,topsep=1pt]
    \item In \Cref{sec:ml-100k-all-archs}, we conduct experiments for all the five recommender architectures (matrix factorization, auto-encoders, neural collaborative filtering, factorization machine, and deep factorization machines) trained using the \underline{\textsc{Movielens-100K}} dataset. The goal is to demonstrate the effectiveness of \toolname to provide explanations across a wide variety of CF models. From the results, we conclude that \toolname indeed performs well on all metrics across the five architectures, demonstrating its generalizability. 

    \item In \Cref{sec:removal-based}), we provide a comparison of one feature removal and SHAP when used for feature attribution by \toolname. 
    \toolname performs marginally better when using SHAP, however, SHAP is \textcolor{blue}{70} times more expensive than one feature removal technique. Therefore, in the evaluation section, we used one feature removal. 

    \item In \Cref{sec:ablation-expt-plots}, we plot the change in metrics as the number of users who rated small number of items (below a threshold) are removed from the dataset, and the recommender system and the auxiliary model are retrained on this dataset. The goal of this experiment is to demonstrate the effectiveness of \toolname across different rating sparsity thresholds. \toolname performs consistently well across the thresholds. 
\end{enumerate}

In \Cref{sec:case-studies}, we dive into two case studies demonstrating the explanations \toolname provides for two real users from the \underline{\textsc{Movielens-100K}} dataset. 

\section{Limitations and Conclusions}
\label{sec:conclusion}

In this work, we proposed a novel attribute-based explainability technique, \toolname, which explains recommendations for collaborative filtering-based recommender systems using the attributes of the item. 
To the best of our knowledge, \toolname is the \textbf{first and only} existing technique that provides such explanations in a post-hoc and amortized manner. 
We studied the performance of \toolname and \baseline baselines using five large scale recommender systems datasets trained on a variety of collaborative filtering architectures. 
As our results indicate, \toolname strikes an excellent balance between coverage and personalization of the explanations and performs better than all the \baseline baselines on most metrics. 

The minor limitation of \toolname is the cost of training of the auxiliary model. For all five recommender systems, training of the auxiliary model took between 10-30 minutes, depending on the dataset size (significantly less than training the recommender systems which took 6-12 hours). Once trained, these models can be used to explain recommendations for any item for any user. The explanation generation during inference is very cheap; it took less than 10 minutes to compute the general and specific preferences for all the users, even for the largest datasets. This is significantly faster in contrast to LIME-RS which would take 11 days and AMCF-PH which took 1-3 hours. 



\bibliography{refs}

\begin{thebibliography}{50}
\providecommand{\natexlab}[1]{#1}

\bibitem[{Abdollahi and Nasraoui(2017)}]{Adbollahi-and-Nasraoui-constrain}
Abdollahi, B.; and Nasraoui, O. 2017.
\newblock Using Explainability for Constrained Matrix Factorization.
\newblock In \emph{Proceedings of the Eleventh ACM Conference on Recommender
  Systems}, RecSys '17, 79–83. New York, NY, USA: Association for Computing
  Machinery.
\newblock ISBN 9781450346528.

\bibitem[{Adomavicius and Tuzhilin(2005)}]{Adomavicius2005Toward}
Adomavicius, G.; and Tuzhilin, A. 2005.
\newblock {Toward the Next Generation of Recommender Systems: A Survey of the
  State-of-the-Art and Possible Extensions}.
\newblock \emph{IEEE Transactions on Knowledge and Data Engineering}, 17(6):
  734--749.

\bibitem[{Aggarwal(2016)}]{content2}
Aggarwal, C.~C. 2016.
\newblock Content-based recommender systems.
\newblock In \emph{Recommender systems}, 139--166. Springer.

\bibitem[{Bauman, Liu, and Tuzhilin(2017)}]{Bauman2017}
Bauman, K.; Liu, B.; and Tuzhilin, A. 2017.
\newblock Aspect Based Recommendations: Recommending Items with the Most
  Valuable Aspects Based on User Reviews.
\newblock In \emph{Proceedings of the 23rd ACM SIGKDD International Conference
  on Knowledge Discovery and Data Mining}, KDD '17. New York, NY, USA:
  Association for Computing Machinery.
\newblock ISBN 9781450348874.

\bibitem[{Bilgic and Mooney(2005)}]{xai-reco-motivation2}
Bilgic, M.; and Mooney, R. 2005.
\newblock Explaining Recommendations: Satisfaction vs. Promotion.
\newblock In \emph{Proceedings of Beyond Personalization 2005: A Workshop on
  the Next Stage of Recommender Systems Research at the 2005 International
  Conference on Intelligent User Interfaces}.

\bibitem[{Bobadilla et~al.(2013)Bobadilla, Ortega, Hernando, and
  Guti{\'e}rrez}]{recobroad1}
Bobadilla, J.; Ortega, F.; Hernando, A.; and Guti{\'e}rrez, A. 2013.
\newblock Recommender systems survey.
\newblock \emph{Knowledge-based systems}, 46: 109--132.

\bibitem[{Burke(2004)}]{Burke2004HybridRS}
Burke, R. 2004.
\newblock Hybrid Recommender Systems: Survey and Experiments.
\newblock \emph{User Modeling and User-Adapted Interaction}, 12: 331--370.

\bibitem[{Candillier et~al.(2009)Candillier, Jack, Fessant, and
  Meyer}]{content1}
Candillier, L.; Jack, K.; Fessant, F.; and Meyer, F. 2009.
\newblock State-of-the-art recommender systems.
\newblock In \emph{Collaborative and Social Information Retrieval and Access:
  Techniques for Improved User Modeling}, 1--22. IGI Global.

\bibitem[{Chen et~al.(2018)Chen, Zhuang, Hong, Ao, Xie, and He}]{Chen2018}
Chen, J.; Zhuang, F.; Hong, X.; Ao, X.; Xie, X.; and He, Q. 2018.
\newblock Attention-Driven Factor Model for Explainable Personalized
  Recommendation.
\newblock In \emph{The 41st International ACM SIGIR Conference on Research and
  Development in Information Retrieval}, SIGIR '18. New York, NY, USA:
  Association for Computing Machinery.

\bibitem[{Chen et~al.(2016)Chen, Qin, Zhang, and Xu}]{Chen2016}
Chen, X.; Qin, Z.; Zhang, Y.; and Xu, T. 2016.
\newblock Learning to Rank Features for Recommendation over Multiple
  Categories.
\newblock In \emph{Proceedings of the 39th International ACM SIGIR Conference
  on Research and Development in Information Retrieval}, SIGIR '16. New York,
  NY, USA: Association for Computing Machinery.
\newblock ISBN 9781450340694.

\bibitem[{Cheng et~al.(2019)Cheng, Shen, Huang, and Zhu}]{FIA-paper-Cheng2019}
Cheng, W.; Shen, Y.; Huang, L.; and Zhu, Y. 2019.
\newblock Incorporating Interpretability into Latent Factor Models via Fast
  Influence Analysis.
\newblock In \emph{Proceedings of the 25th ACM SIGKDD International Conference
  on Knowledge Discovery and Data Mining}, KDD '19. New York, NY, USA:
  Association for Computing Machinery.
\newblock ISBN 9781450362016.

\bibitem[{Covert, Lundberg, and Lee(2021)}]{Ians-paper-xai-by-removal-survey}
Covert, I.; Lundberg, S.; and Lee, S.-I. 2021.
\newblock Explaining by Removing: A Unified Framework for Model Explanation.
\newblock \emph{Journal of Machine Learning Research}, 22(209): 1--90.

\bibitem[{Dror et~al.(2011)Dror, Koenigstein, Koren, and
  Weimer}]{yahoomusicdataset}
Dror, G.; Koenigstein, N.; Koren, Y.; and Weimer, M. 2011.
\newblock The Yahoo! Music Dataset and KDD-Cup'11.
\newblock In \emph{Proceedings of the 2011 International Conference on KDD Cup
  2011 - Volume 18}, KDDCUP'11, 3--18. JMLR.org.

\bibitem[{Ferwerda, Swelsen, and Yang(2018)}]{attribute-based-xai-reco1}
Ferwerda, B.; Swelsen, K.; and Yang, E. 2018.
\newblock Explaining Content-Based Recommendations.
\newblock \emph{-}.

\bibitem[{Google(2022)}]{google-cf-serendipitious}
Google. 2022.
\newblock Collaborative Filtering.
\newblock [Online; accessed 25-September-2022].

\bibitem[{Harper and Konstan(2011)}]{grouplensHetRec2011}
Harper, F.~M.; and Konstan, J.~A. 2011.
\newblock {H}et{R}ec 2011 --- grouplens.org.
\newblock \url{https://grouplens.org/datasets/hetrec-2011/}.
\newblock [Accessed 29-May-2023].

\bibitem[{Harper and Konstan(2015)}]{movielens-dataset}
Harper, F.~M.; and Konstan, J.~A. 2015.
\newblock The MovieLens Datasets: History and Context.
\newblock \emph{ACM Trans. Interact. Intell. Syst.}, 5(4).

\bibitem[{Harper and Konstan(2016)}]{grouplensMovieLens20m}
Harper, F.~M.; and Konstan, J.~A. 2016.
\newblock {M}ovie{L}ens 20{M} {D}ataset --- grouplens.org.
\newblock \url{https://grouplens.org/datasets/movielens/20m/}.
\newblock [Accessed 29-May-2023].

\bibitem[{He et~al.(2015)He, Chen, Kan, and
  Chen}]{attribute-based-xai-reco4-trirank-paper}
He, X.; Chen, T.; Kan, M.-Y.; and Chen, X. 2015.
\newblock TriRank: Review-Aware Explainable Recommendation by Modeling Aspects.
\newblock In \emph{Proceedings of the 24th ACM International on Conference on
  Information and Knowledge Management}, CIKM '15. New York, NY, USA:
  Association for Computing Machinery.
\newblock ISBN 9781450337946.

\bibitem[{Hou et~al.(2019)Hou, Yang, Wu, and
  Yu}]{attribute-based-xai-reco2-radarchart-paper}
Hou, Y.; Yang, N.; Wu, Y.; and Yu, P.~S. 2019.
\newblock Explainable Recommendation with Fusion of Aspect Information.
\newblock \emph{World Wide Web}, 22(1).

\bibitem[{Kaggle(2016)}]{kaggleAnimeRecommendations}
Kaggle. 2016.
\newblock {A}nime {R}ecommendations {D}atabase --- kaggle.com.
\newblock
  \url{https://www.kaggle.com/datasets/CooperUnion/anime-recommendations-database}.
\newblock [Accessed 29-May-2023].

\bibitem[{Lawton(2017)}]{pandora-content-based}
Lawton, G. 2017.
\newblock How Pandora built a better recommendation engine.
\newblock [Online; accessed 25-September-2022].

\bibitem[{Liu et~al.(2019)Liu, Wen, Jing, Yu, Zhang, and Zhang}]{Liu2019}
Liu, H.; Wen, J.; Jing, L.; Yu, J.; Zhang, X.; and Zhang, M. 2019.
\newblock In2Rec: Influence-Based Interpretable Recommendation.
\newblock In \emph{Proceedings of the 28th ACM International Conference on
  Information and Knowledge Management}, CIKM '19. New York, NY, USA:
  Association for Computing Machinery.
\newblock ISBN 9781450369763.

\bibitem[{Lundberg and Lee(2017)}]{shap-paper}
Lundberg, S.~M.; and Lee, S.-I. 2017.
\newblock A Unified Approach to Interpreting Model Predictions.
\newblock In \emph{Proceedings of the 31st NeurIPS}. Red Hook, NY, USA: Curran
  Associates Inc.
\newblock ISBN 9781510860964.

\bibitem[{McAuley and Leskovec(2013)}]{McAuley-and-Leskovec2013}
McAuley, J.; and Leskovec, J. 2013.
\newblock Hidden Factors and Hidden Topics: Understanding Rating Dimensions
  with Review Text.
\newblock In \emph{Proceedings of the 7th ACM Conference on Recommender
  Systems}, RecSys '13. New York, NY, USA: Association for Computing Machinery.
\newblock ISBN 9781450324090.

\bibitem[{McAuley, Leskovec, and
  Jurafsky(2012)}]{McAuley2012LearningAttributes}
McAuley, J.; Leskovec, J.; and Jurafsky, D. 2012.
\newblock Learning Attitudes and Attributes from Multi-aspect Reviews.
\newblock \emph{2012 IEEE 12th International Conference on Data Mining},
  1020--1025.

\bibitem[{Molnar(2022)}]{molnar2022-xaibook}
Molnar, C. 2022.
\newblock \emph{Interpretable Machine Learning}.
\newblock -, 2 edition.

\bibitem[{N\'{o}brega and Marinho(2019)}]{limers-paper}
N\'{o}brega, C.; and Marinho, L. 2019.
\newblock Towards Explaining Recommendations through Local Surrogate Models.
\newblock In \emph{Proceedings of the 34th ACM/SIGAPP Symposium on Applied
  Computing}, SAC '19. New York, NY, USA: Association for Computing Machinery.
\newblock ISBN 9781450359337.

\bibitem[{Pan et~al.(2020)Pan, Li, Li, and Zhu}]{amcf-paper}
Pan, D.; Li, X.; Li, X.; and Zhu, D. 2020.
\newblock Explainable Recommendation via Interpretable Feature Mapping and
  Evaluation of Explainability.
\newblock In \emph{Proceedings of the Twenty-Ninth International Joint
  Conference on Artificial Intelligence, {IJCAI-20}}, IJCAI'20, 2690--2696.
  International Joint Conferences on Artificial Intelligence Organization.
\newblock Main track.

\bibitem[{Peake and Wang(2018)}]{Peake-and-Wang-data-mining2018}
Peake, G.; and Wang, J. 2018.
\newblock Explanation Mining: Post Hoc Interpretability of Latent Factor Models
  for Recommendation Systems.
\newblock In \emph{Proceedings of the 24th ACM SIGKDD International Conference
  on Knowledge Discovery and Data Mining}, KDD '18, 2060–2069. New York, NY,
  USA: Association for Computing Machinery.
\newblock ISBN 9781450355520.

\bibitem[{Ribeiro, Singh, and Guestrin(2016)}]{ribeiro_why_2016}
Ribeiro, M.~T.; Singh, S.; and Guestrin, C. 2016.
\newblock "Why Should I Trust You?": Explaining the Predictions of Any
  Classifier.
\newblock In \emph{Proceedings of the 22nd ACM SIGKDD International Conference
  on Knowledge Discovery and Data Mining}, KDD '16. New York, NY, USA:
  Association for Computing Machinery.
\newblock ISBN 9781450342322.

\bibitem[{Roy(2020)}]{collab-reco}
Roy, A. 2020.
\newblock Introduction To Recommender Systems- 1: Content-Based Filtering And
  Collaborative Filtering.
\newblock
  \url{https://towardsdatascience.com/introduction-to-recommender-systems-1-971bd274f421}.
\newblock [Online; accessed 28-December-2021].

\bibitem[{Shapley(1953)}]{shapley:book1952}
Shapley, L.~S. 1953.
\newblock A Value for n-Person Games.
\newblock In Kuhn, H.~W.; and Tucker, A.~W., eds., \emph{Contributions to the
  Theory of Games II}, 307--317. Princeton: Princeton University Press.

\bibitem[{Sinha and Swearingen(2002)}]{old-xai-reco-paper-sinha-2002}
Sinha, R.; and Swearingen, K. 2002.
\newblock The Role of Transparency in Recommender Systems.
\newblock In \emph{CHI '02 Extended Abstracts on Human Factors in Computing
  Systems}, CHI EA '02, 830–831. New York, NY, USA: Association for Computing
  Machinery.
\newblock ISBN 1581134541.

\bibitem[{Su and Khoshgoftaar(2009)}]{colab1}
Su, X.; and Khoshgoftaar, T.~M. 2009.
\newblock A survey of collaborative filtering techniques.
\newblock \emph{Advances in artificial intelligence}, 2009.

\bibitem[{Tao et~al.(2019)Tao, Jia, Wang, and Wang}]{Tao2019}
Tao, Y.; Jia, Y.; Wang, N.; and Wang, H. 2019.
\newblock The FacT: Taming Latent Factor Models for Explainability with
  Factorization Trees.
\newblock In \emph{Proceedings of the 42nd International ACM SIGIR Conference
  on Research and Development in Information Retrieval}, SIGIR'19. New York,
  NY, USA: Association for Computing Machinery.
\newblock ISBN 9781450361729.

\bibitem[{Tintarev(2007)}]{Tintarev2007-thesis}
Tintarev, N. 2007.
\newblock Explanations of Recommendations.
\newblock In \emph{Proceedings of the 2007 ACM Conference on Recommender
  Systems}, RecSys '07, 203–206. New York, NY, USA: Association for Computing
  Machinery.
\newblock ISBN 9781595937308.

\bibitem[{Tintarev and Masthoff(2007)}]{Tintarev-cfe-xai-survey-2007}
Tintarev, N.; and Masthoff, J. 2007.
\newblock A Survey of Explanations in Recommender Systems.
\newblock In \emph{2007 IEEE 23rd International Conference on Data Engineering
  Workshop}, 801--810.

\bibitem[{Tintarev and Masthoff(2011)}]{Tintarev2011}
Tintarev, N.; and Masthoff, J. 2011.
\newblock \emph{Designing and Evaluating Explanations for Recommender Systems},
  479--510.
\newblock Boston, MA: Springer US.

\bibitem[{Varshney(2021)}]{varshney}
Varshney, K.~R. 2021.
\newblock \emph{Trustworthy Machine Learning}.
\newblock Chappaqua, NY, USA: -.
\newblock Http://trustworthymachinelearning.com.

\bibitem[{Verma, Hines, and Dickerson(2021)}]{verma2021amortized}
Verma, S.; Hines, K.; and Dickerson, J.~P. 2021.
\newblock Amortized Generation of Sequential Counterfactual Explanations for
  Black-box Models.

\bibitem[{Vig, Sen, and
  Riedl(2009)}]{attribute-based-xai-reco3-tagplanations-paper}
Vig, J.; Sen, S.; and Riedl, J. 2009.
\newblock Tagsplanations: Explaining Recommendations Using Tags.
\newblock In \emph{Proceedings of the 14th International Conference on
  Intelligent User Interfaces}, IUI '09. New York, NY, USA: Association for
  Computing Machinery.
\newblock ISBN 9781605581682.

\bibitem[{Wang et~al.(2018)Wang, Chen, Yang, Wu, Wu, and
  Xie}]{rl-approach-xai-reco-MSFT}
Wang, X.; Chen, Y.; Yang, J.; Wu, L.; Wu, Z.; and Xie, X. 2018.
\newblock A Reinforcement Learning Framework for Explainable Recommendation.
\newblock In \emph{2018 IEEE International Conference on Data Mining (ICDM)},
  587--596.

\bibitem[{Wang et~al.(2022)Wang, Li, Yu, and Xu}]{attribute-based-cfe-reco-RL}
Wang, X.; Li, Q.; Yu, D.; and Xu, G. 2022.
\newblock Reinforced Path Reasoning for Counterfactual Explainable
  Recommendation.

\bibitem[{Webber, Moffat, and Zobel(2010)}]{RBO-paper}
Webber, W.; Moffat, A.; and Zobel, J. 2010.
\newblock A Similarity Measure for Indefinite Rankings.
\newblock \emph{ACM Trans. Inf. Syst.}, 28(4).

\bibitem[{{Wikipedia contributors}(2022)}]{rank-correlation}
{Wikipedia contributors}. 2022.
\newblock Rank correlation --- {Wikipedia}{,} The Free Encyclopedia.
\newblock [Online; accessed 25-September-2022].

\bibitem[{Woollacott(2018)}]{netflix-no-more-reviews}
Woollacott, E. 2018.
\newblock Why Netflix Won't Let You Write Reviews Any More.
\newblock [Online; accessed 25-September-2022].

\bibitem[{Xian et~al.(2021)Xian, Zhao, Li, Chan, Kan, Ma, Dong, Faloutsos,
  Karypis, Muthukrishnan, and Zhang}]{EX3-yongfeng-paper}
Xian, Y.; Zhao, T.; Li, J.; Chan, J.; Kan, A.; Ma, J.; Dong, X.~L.; Faloutsos,
  C.; Karypis, G.; Muthukrishnan, S.; and Zhang, Y. 2021.
\newblock EX3: Explainable Attribute-Aware Item-Set Recommendations.
\newblock In \emph{Proceedings of the 15th ACM Conference on Recommender
  Systems}, RecSys '21. New York, NY, USA: Association for Computing Machinery.

\bibitem[{Zhang and Chen(2020)}]{survey-reco-xai}
Zhang, Y.; and Chen, X. 2020.
\newblock Explainable Recommendation: A Survey and New Perspectives.
\newblock \emph{Found. Trends Inf. Retr.}, 14: 1--101.

\bibitem[{Zhang et~al.(2014)Zhang, Lai, Zhang, Zhang, Liu, and
  Ma}]{explicit-factors-paper-yongfeng}
Zhang, Y.; Lai, G.; Zhang, M.; Zhang, Y.; Liu, Y.; and Ma, S. 2014.
\newblock Explicit Factor Models for Explainable Recommendation Based on
  Phrase-Level Sentiment Analysis.
\newblock In \emph{Proceedings of the 37th International ACM SIGIR Conference
  on Research and Development in Information Retrieval}, SIGIR '14. New York,
  NY, USA: Association for Computing Machinery.
\newblock ISBN 9781450322577.

\end{thebibliography}

\clearpage

\appendix

\section{Dataset details}
\label{sec:datasetdetails}
We provide details of the five datasets we used in the evaluation section. Each of these datasets were used for training different recommender systems based on various architectures (see \Cref{tab:best-test-loss} for the architecture chosen for each dataset). 

\begin{enumerate}[leftmargin=*,noitemsep,topsep=-1pt]
    \item \underline{\textsc{Movielens-$100K$}} \citep{movielens-dataset} dataset is trained using \emph{Matrix Factorization} model. This dataset has 100,000 ratings for 1,700 movies provided by 1,000 users. Each rating is an integer from 1 to 5, with a higher rating meaning a higher likeness for the movie. The dataset also contains the genre of each movie, and there are a total of 18 genres in the dataset: \emph{Action, Adventure, Animation, Children's, Comedy, Crime, Documentary, Drama, Fantasy, Film-Noir, Horror, Musical, Mystery, Romance, Sci-Fi, Thriller, War, Western}. Each movie usually belongs to 1-3 genres. 

    \item \underline{\textsc{Hetrec}} \citep{grouplensHetRec2011} dataset is trained using \emph{Matrix Factorization} model. This dataset has 855,000 ratings for 10,100 movies provided by 2,100 users. Similar to \emph{Movielens-$100K$}, this dataset has integer ratings from 1 to 5, and each movie has genre annotations. There are 20 genres in the dataset; it shares the 18 with \emph{Movielens-$100K$} and has two additional genres: \emph{IMAX} and \emph{Short}. 

    \item \underline{\textsc{Anime}} \citep{kaggleAnimeRecommendations} dataset is trained using \emph{Neural Collaborative Filtering} model. This dataset has 6.1 million ratings for 6,500 animes provided by 47,000 users. Each rating is an integer from 1 to 10. Each anime belongs to 1-4 genres from a total set of 49 genres. 

    \item \underline{\textsc{Movielens-$20M$}} \citep{grouplensMovieLens20m} dataset is trained using \emph{Deep Factorization Machine}. This dataset has 20 million ratings for 27,000 movies provided by 138,000 users. Similar to \emph{Movielens-$100K$}, this dataset has integer ratings from 1-5, and each movie has genre annotations. There are 19 total genres in the dataset; it shares the 18 with \emph{Movielens-$100K$} and has one additional genre: \emph{IMAX}.

    \item \underline{\textsc{YahooMusic}} \citep{yahoomusicdataset} dataset is trained using \emph{Matrix Factorization} model. This dataset has 32.6 million ratings for 128,000 tracks or albums provided by 75,000 users. Each rating is an integer between 0 and 10. Each track or album has 1-5 genre annotations from a total set of 41 genres. 

\end{enumerate}

\label{sec:appendix}
\section{Related Work}
\label{sec:related}
In this section we provide a brief review of relevant works that cover the general topic of explainability in recommender systems, as well as post-hoc explainability and attribute-based explainability as specific topics. 

\subsection{Explainability in Recommender Systems}
Explainable recommender systems provide recommendations along with an explanation for doing so. The term `explainable recommendation' was introduced by \citet{explicit-factors-paper-yongfeng} in \citeyear{explicit-factors-paper-yongfeng}; however, papers were talking about the benefits of providing explanations much earlier \citep{Tintarev-cfe-xai-survey-2007,Tintarev2011,old-xai-reco-paper-sinha-2002}. 

Most previous methods for explainability in collaborative filtering-based (CF-based) recommender systems provide explanations in user-based or item-based fashion \citep{survey-reco-xai}. 
The approaches that generate these explanations can either be model-specific or model agnostic, the former kind constituting the majority of previous approaches. 
Models-specific approaches train inherently interpretable models by either constraining the embedding space or developing custom recommender system architectures that are not usually generalizable. For example, \citet{explicit-factors-paper-yongfeng} propose extracting user preferences over attributes from their reviews and using that to provide explanations for recommendations. The authors train an explicit factor recommender model that takes the user's preference over attributes as input during training the recommender model. Such an approach is restricted to situations where the user reviews over items are available in abundance and also might suffer from the accuracy-interpretability trade-off. 
In another approach, \citet{Adbollahi-and-Nasraoui-constrain} add constraints to the latent space of the matrix factorization models to encourage explainability. Even though their approach only provides user and item-based explanations, it lands deep in the accuracy-interpretability trade-off debate. 
There exist many other approaches in this category \citep{Chen2016,Chen2018,Bauman2017,Tao2019,Liu2019,McAuley-and-Leskovec2013,EX3-yongfeng-paper}. 
As evident, the major downside of these approaches is that they are \emph{not} post-hoc, i.e., they need to modify the recommender system architecture in order to provide explanations -- and this is problematic for two reasons: 1) there is a potential accuracy-interpretability trade-off, and 2) real-world recommender systems are trained and fine-tuned over large datasets, and their developers are rarely willing to train a new model from scratch in order to provide explanations \citep{varshney}. 
Hence post-hoc explainability techniques attract a lot of attention, especially from an industry perspective \citep{verma2021amortized,molnar2022-xaibook}. 

\subsection{Post-hoc Explainability in Recommender Systems}
There only exist a handful of explainability techniques in this category. 
\citet{Peake-and-Wang-data-mining2018} use a data-mining approach to provide item-based explanations in a post-hoc manner. 
\citet{limers-paper} propose LIME-RS which draws motivation from the original LIME paper \citep{ribeiro_why_2016}. LIME-RS samples items close to a recommended item, and learns a linear regression model on these samples, and uses the regressor to provide an explanation that can be either item-based or attribute-based (the latter can be provided if the attributes of the items are appended to the items embeddings when training the regression model). LIME-RS is the only previous approach that provides attribute-based explanations for CF-based recommender systems and is a post-hoc explanation technique. However, LIME-RS is not amortized -- it trains a separate regression model for each recommended item and hence is not scalable to large scale datasets (see \Cref{sec:eval} for evaluation of LIME-RS on large scale datasets). 
\citet{FIA-paper-Cheng2019} propose a technique that uses influence functions to find the influence of training ratings on a particular recommendation, and the ratings with the greatest influence are served as explanations. Hence, this technique also only provides item-based explanations. 

\subsection{Attribute-Based Explainability in Recommender Systems}
Unlike user-based and item-based explanations, attribute-based explanations explain a recommendation based on the user's personalized attribute preferences, thereby enhancing the system's overall persuasiveness and trustworthiness \citep{attribute-based-cfe-reco-RL,attribute-based-xai-reco1}. 
As mentioned earlier, there do exist previous works that provide explanations that utilize the user's preference over interpretable attributes of an item. Most of these works utilize the reviews provided by the users to capture this preference (and use that to train the recommender system itself \citep{explicit-factors-paper-yongfeng,McAuley-and-Leskovec2013,attribute-based-xai-reco2-radarchart-paper}). However, such reviews might not exist even in the presence of ratings. For example, Netflix scrapped its review section due to low user participation recently \citep{netflix-no-more-reviews}. 
Secondly, there is no guarantee of mention of interpretable attributes in the reviews that can be used to learn the user's preference. Even the datasets used in these approaches had very sparse user reviews; over 77\% of the users had only \emph{one review} \citep{attribute-based-xai-reco4-trirank-paper}, making the inference of users' preferences challenging. 

On the other hand, \toolname also provides attribute-based explanations; however, it utilizes the attributes of an item that is available in their metadata, e.g., movie or game genres, and hence \emph{does not} depend on user reviews. 
Similar to ours, AMCF \citep{amcf-paper} learns users' preferences over item attributes by training an attention network. So it provides attribute-based explanations like our approach; however, AMCF is not a post-hoc technique -- it needs to be trained along with the recommender system. 
Similarly, \citet{attribute-based-cfe-reco-RL} proposes CERec to determine the attributes that are important for a user-item pair (in a counterfactual manner); however, CERec is also not a post-hoc technique. 
\citet{rl-approach-xai-reco-MSFT} propose a post-hoc RL-based approach that generates attribute-based explanations for a user; however, like most previous approaches, it acquires the attributes from the user reviews -- whose downsides are aforementioned. 
Hence our approach sits at the intersection of attribute-based (\emph{without} requiring user reviews), post-hoc explanation techniques, generating amortized explanations. 
To the best of our knowledge, \toolname \textbf{is the first technique} to have all these desirable properties. 


\section{Attribute Distribution Plots}
\label{sec:attribute-dist-plots}

\begin{figure}[h]
        \centering
        \includegraphics[width=0.8\columnwidth]{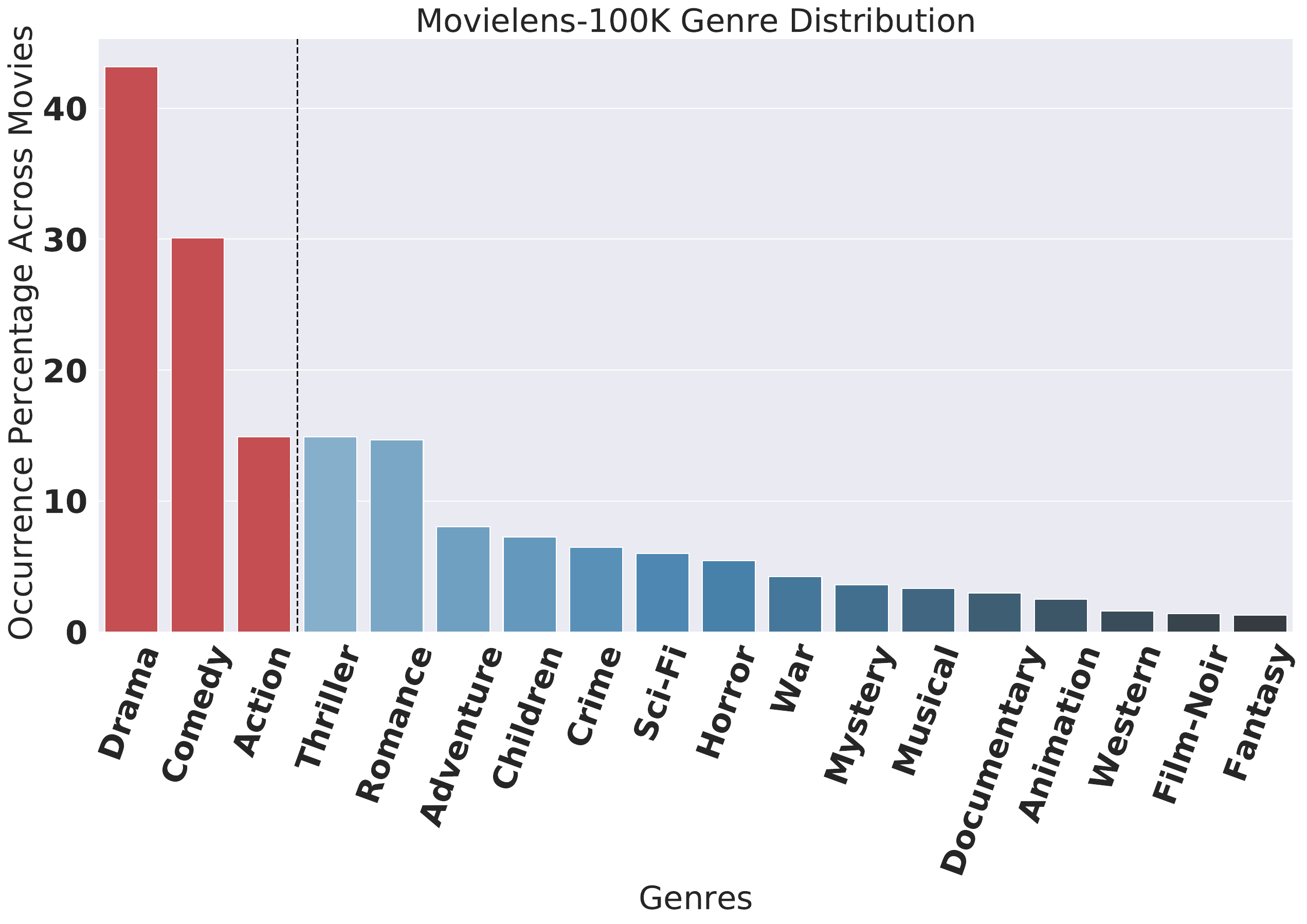}
        \caption{The genre distribution for \underline{\textsc{Movielens-$100K$}} dataset. The top-3 most popular genres: \emph{Action}, \emph{Comedy}, and \emph{Drama} occur in over 88\% of the movies. And therefore, global popularity can achieve very high test set coverage while providing uninformative and unpersonalized explanations.}
        \label{fig:genre_dist_ml-100k-repeat}
    \end{figure}

\begin{figure}
        \centering
        \includegraphics[width=0.9\columnwidth]{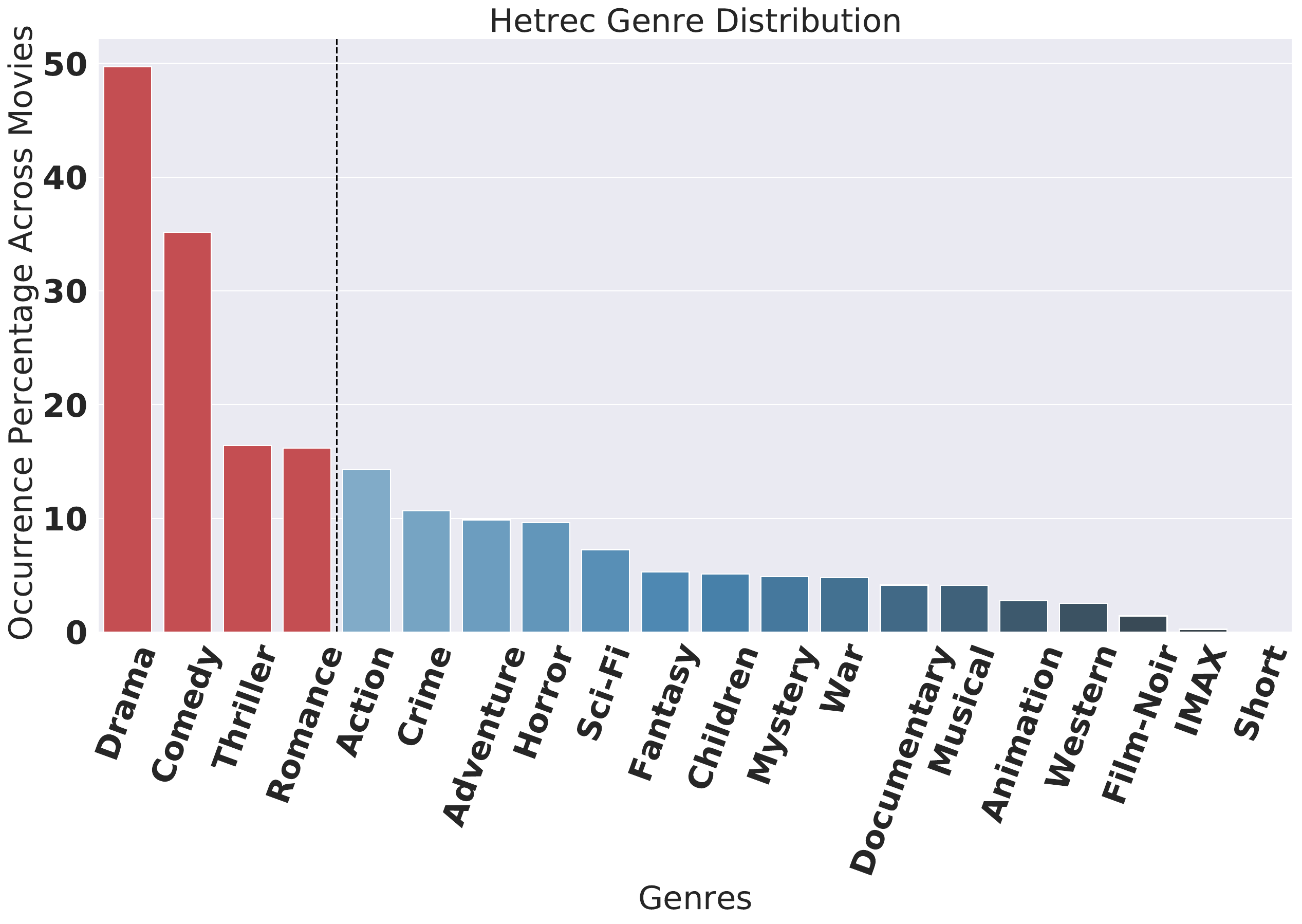}
        \caption{The genre distribution for \underline{\textsc{Hetrec}} dataset. The top-4 most popular genres: \emph{Comedy}, \emph{Drama}, \emph{Romance}, and \emph{Thriller} occur in over 95\% of the movies. And therefore, global popularity can achieve very high test set coverage while providing uninformative and non-personalized explanations. }
        \label{fig:genre_dist_hetrec}
\end{figure}

Here we plot the distribution of genres for \underline{\textsc{Movielens-$100K$}} (\Cref{fig:genre_dist_ml-100k-repeat})and \underline{\textsc{Hetrec}} (\Cref{fig:genre_dist_hetrec}) datasets. There is a large skew in the distribution, with the top-3 genres and top-4 genres occurring in more than 88\% and 95\% of their items respectively. 

\section{Case Studies}
\label{sec:case-studies}

In this section we dive into two case studies demonstrating the use of \toolname. 

\textbf{A.  A case study where the popularity of a genre is insufficient to understand user preferences}:

A user has liked 4 and disliked 23 movies in the \underline{\textsc{Movielens-100K}} dataset. The liked movies and their respective genres are:
\begin{enumerate}
    \item Good Will Hunting: \textit{Drama}
    \item Apt Pupil: \textit{Drama, Thriller}
    \item Fast, Cheap \& Out of Control: \textit{Documentary}
    \item Welcome To Sarajevo: \textit{Drama, War}
\end{enumerate}

The user popularity identifies \textit{Drama} to be the most preferred attribute for this user. However, \toolname identifies that \textit{Documentary} is more important to this user.
Indeed, \toolname's identification is accurate as 7 out of the 23 disliked items also belong to the \textit{Drama} genre.

\textbf{B.  A case study where \toolname explains the top movie recommendations better than popularity}:

A user has liked 4 and disliked 10 movies in the \underline{\textsc{Movielens-100K}} dataset, and their liked movies and their respective genres are:
\begin{enumerate}
    \item Ulee's Gold: \textit{Drama}
    \item Everyone Says I Love You: \textit{Comedy, Musical, Romance}
    \item Wag the Dog: \textit{Comedy, Drama}
    \item Kundun: \textit{Drama}
\end{enumerate}

Among the movie attributes, we can see that \textit{Drama} is more popular than \textit{Comedy}, however \toolname identifies that the top genre preferences of this user is \textit{Comedy}. 

Looking at the top-3 new movie recommendations generated by the recommender system for the user unveils \toolname's capability:
\begin{enumerate}
    \item Casablanca: \textit{Drama, Romance, War}
    \item A Close Shave: \textit{Animation, Comedy, Thriller}
    \item The Wrong Trousers: \textit{Animation, Comedy}
\end{enumerate}

Among the top-3 recommendations, \textit{Drama} only occurs in 1, whereas \textit{Comedy} occurs in 2. Thereby illustrating the effectiveness of \toolname's explanations. 

\section{Evaluation Metrics for \underline{\textsc{Anime}} and \underline{\textsc{YahooMusic}} datasets}
\label{sec:remainingdatasets}
\begin{table*}[h]
	\caption{The test set coverage, top-k recommendations coverage, and the 8 explanation personalization metrics are reported here (averaged over all users) for the \underline{\textsc{Neural CF}} trained using \underline{\textsc{Anime}} dataset. The mean and standard deviation value for each metric is computed over 5 random data splits. We compare 3 architecture choices for auxiliary modeling of RecXplainer. The best results are highlighted in bold. }
	\begin{center}
	\resizebox{1.0\textwidth}{!}{%
		\begin{tabular}{lllllllll}
			\toprule
			Metrics & LIME-RS & AMCF-PH & GBL Popl. & User Popl. & Random & RX-Linear & RX-MLP & RX-GBDT \\
			\midrule
			Testset Coverage & \textit{Timeout} & 64.55 $\pm$ 4.27 & \textbf{91.1 $\pm$ 0.01} & 89.84 $\pm$ 0.02 & 59.36 $\pm$ 0.11 & 74.13 $\pm$ 2.23 & 81.52 $\pm$ 1.36 & 83.35 $\pm$ 0.13 \\
			Recommendations Coverage & \textit{Timeout} & 53.85 $\pm$ 3.81 & \textbf{71.49 $\pm$ 1.85} & \textbf{69.68 $\pm$ 1.55} & 50.16 $\pm$ 0.45 & \textbf{69.99 $\pm$ 1.62} & \textbf{72.67 $\pm$ 1.95} & \textbf{71.22 $\pm$ 0.8} \\
			\midrule
			CondProb Generalpref Coverage & \textit{Timeout} & 61.56 $\pm$ 3.68 & 40.74 $\pm$ 0.17 & 43.94 $\pm$ 0.24 & 62.46 $\pm$ 0.17 & \textbf{68.54 $\pm$ 2.72} & 57.76 $\pm$ 1.0 & 61.96 $\pm$ 0.35 \\
			CondProb Generalpref Ranking & \textit{Timeout} & 11.82 $\pm$ 1.72 & 4.84 $\pm$ 0.03 & 6.98 $\pm$ 0.04 & 12.5 $\pm$ 0.08 & \textbf{18.02 $\pm$ 0.78} & 12.41 $\pm$ 0.23 & 13.25 $\pm$ 0.14 \\
			CondProb Specificpref Coverage & \textit{Timeout} & 64.7 $\pm$ 0.49 & 40.74 $\pm$ 0.17 & 43.94 $\pm$ 0.24 & 62.45 $\pm$ 0.02 & \textbf{66.76 $\pm$ 0.36} & 63.11 $\pm$ 0.49 & 63.41 $\pm$ 0.08 \\
			CondProb Specificpref Ranking & \textit{Timeout} & 40.73 $\pm$ 0.95 & 3.4 $\pm$ 0.02 & 5.51 $\pm$ 0.03 & 10.53 $\pm$ 0.0 & \textbf{47.17 $\pm$ 0.26} & 42.42 $\pm$ 0.07 & 45.11 $\pm$ 0.12 \\
			Odds Generalpref Coverage & \textit{Timeout} & 66.29 $\pm$ 5.69 & 88.26 $\pm$ 0.09 & 88.71 $\pm$ 0.08 & 62.31 $\pm$ 0.21 & 88.61 $\pm$ 2.53 & 91.78 $\pm$ 0.43 & \textbf{93.83 $\pm$ 0.14} \\
			Odds Generalpref Ranking & \textit{Timeout} & 15.99 $\pm$ 3.94 & 41.67 $\pm$ 0.05 & \textbf{70.13 $\pm$ 0.06} & 12.5 $\pm$ 0.06 & 24.43 $\pm$ 1.73 & 35.54 $\pm$ 1.24 & 41.51 $\pm$ 0.17 \\
			Odds Specificpref Coverage & \textit{Timeout} & 76.01 $\pm$ 1.91 & 88.26 $\pm$ 0.1 & \textbf{88.71 $\pm$ 0.08} & 62.4 $\pm$ 0.05 & 81.57 $\pm$ 0.17 & 81.69 $\pm$ 0.39 & 83.54 $\pm$ 0.01 \\
			Odds Specificpref Ranking & \textit{Timeout} & 41.5 $\pm$ 1.91 & 38.93 $\pm$ 0.05 & \textbf{69.38 $\pm$ 0.07} & 10.52 $\pm$ 0.01 & 36.86 $\pm$ 0.43 & 40.66 $\pm$ 0.39 & 43.35 $\pm$ 0.06 \\
			\bottomrule
		\end{tabular}
	}
	\end{center}
	\label{tab:5-random-seeds-anime-ncf}
\end{table*}

\begin{table*}[h]
	\caption{The test set coverage, top-k recommendations coverage, and the 8 explanation personalization metrics are reported here (averaged over all users) for the \underline{\textsc{Matrix Factorization}} model trained using \underline{\textsc{YahooMusic}} dataset. The mean and standard deviation value for each metric is computed over 5 random data splits. We compare 3 architecture choices for auxiliary modeling of RecXplainer. The best results are highlighted in bold. }
	\begin{center}
	\resizebox{1.0\textwidth}{!}{%
		\begin{tabular}{lllllllll}
			\toprule
			Metrics & LIME-RS & AMCF-PH & GBL Popl. & User Popl. & Random & RX-Linear & RX-MLP & RX-GBDT \\
			\midrule
			Testset Coverage & \textit{Timeout} & 66.31 $\pm$ 0.49 & 94.81 $\pm$ 0.02 & \textbf{97.4 $\pm$ 0.02} & 51.63 $\pm$ 0.04 & 96.63 $\pm$ 0.02 & 95.51 $\pm$ 0.65 & 93.78 $\pm$ 0.03 \\
			Recommendations Coverage & \textit{Timeout} & 61.01 $\pm$ 1.13 & 91.29 $\pm$ 0.17 & \textbf{92.41 $\pm$ 0.14} & 37.24 $\pm$ 0.1 & 91.76 $\pm$ 0.14 & 90.22 $\pm$ 0.67 & 87.15 $\pm$ 0.1 \\
			\midrule
			CondProb Generalpref Coverage & \textit{Timeout} & 94.91 $\pm$ 0.24 & 78.44 $\pm$ 0.06 & 89.39 $\pm$ 0.07 & 78.6 $\pm$ 0.15 & 93.12 $\pm$ 0.05 & 93.47 $\pm$ 0.29 & \textbf{95.37 $\pm$ 0.07} \\
			CondProb Generalpref rankingtop5 & \textit{Timeout} & 27.77 $\pm$ 0.46 & 11.4 $\pm$ 0.01 & 24.73 $\pm$ 0.01 & 14.29 $\pm$ 0.06 & 28.81 $\pm$ 0.03 & 29.16 $\pm$ 0.22 & \textbf{33.58 $\pm$ 0.02} \\
			CondProb Specificpref Coverage & \textit{Timeout} & 86.07 $\pm$ 0.09 & 78.37 $\pm$ 0.06 & \textbf{89.36 $\pm$ 0.07} & 78.54 $\pm$ 0.01 & 86.63 $\pm$ 0.04 & 87.32 $\pm$ 0.02 & 88.19 $\pm$ 0.05 \\
			CondProb Specificpref Ranking & \textit{Timeout} & 52.43 $\pm$ 0.33 & 5.92 $\pm$ 0.02 & 13.26 $\pm$ 0.01 & 10.0 $\pm$ 0.01 & 46.17 $\pm$ 0.02 & 49.9 $\pm$ 0.19 & \textbf{53.91 $\pm$ 0.05} \\
			Odds Generalpref Coverage & \textit{Timeout} & \textbf{95.56 $\pm$ 0.1} & 71.63 $\pm$ 0.11 & 86.28 $\pm$ 0.05 & 78.49 $\pm$ 0.18 & 90.61 $\pm$ 0.07 & 91.53 $\pm$ 0.3 & 93.36 $\pm$ 0.04 \\
			Odds Generalpref rankingtop5 & \textit{Timeout} & 29.35 $\pm$ 0.43 & 11.05 $\pm$ 0.02 & 23.67 $\pm$ 0.02 & 14.29 $\pm$ 0.03 & 27.08 $\pm$ 0.04 & 28.01 $\pm$ 0.22 & \textbf{31.91 $\pm$ 0.03} \\
			Odds Specificpref Coverage & \textit{Timeout} & 84.66 $\pm$ 0.12 & 71.79 $\pm$ 0.11 & \textbf{86.41 $\pm$ 0.04} & 78.54 $\pm$ 0.03 & 83.09 $\pm$ 0.03 & 84.25 $\pm$ 0.05 & 84.9 $\pm$ 0.03 \\
			Odds Specificpref Ranking & \textit{Timeout} & 53.03 $\pm$ 0.16 & 6.97 $\pm$ 0.02 & 14.34 $\pm$ 0.01 & 10.0 $\pm$ 0.01 & 46.21 $\pm$ 0.03 & 50.36 $\pm$ 0.18 & \textbf{54.32 $\pm$ 0.04} \\
			\bottomrule
		\end{tabular}
	}
	\end{center}
	\label{tab:5-random-seeds-yahoomusic-matrix_fact}
\end{table*}

\Cref{tab:5-random-seeds-anime-ncf} and \Cref{tab:5-random-seeds-yahoomusic-matrix_fact} reports the metrics for the \underline{\textsc{Anime}} and \underline{\textsc{YahooMusic}} datasets respectively. We analyze the results below:
\begin{enumerate}[leftmargin=*,noitemsep,topsep=-1pt]
    \item \emph{LIME-RS: } LIME-RS cannot scale to both these datasets (overall, it did not scale for 3 out of 5 datasets). For the Anime dataset, the estimated computation time for specific and general preferences is over 3 days; for the YahooMusic dataset, the estimated computation time is over 18 days. 
    
    \item \emph{AMCF-PH: } \toolname performs better than AMCF-PH for all ten metrics and nine out of ten metrics for Anime and YahooMusic datasets respectively. 
    
    \item \emph{Global popularity: } For Anime dataset, global popularity performs the best on coverage metrics, however, \toolname performs better than it for YahooMusic dataset on those metrics. 
    \toolname performs better than global popularity on seven out of eight and all eight personalization metrics for Anime and YahooMusic datasets respectively. 
    
    \item \emph{User-specific popularity: } User-specific popularity performs the second best for the Anime dataset and the best for YahooMusic dataset on the coverage metrics. \toolname performs better than user-specific popularity on five out of eight and six out of eight personalization metrics for Anime and YahooMusic datasets respectively. 
    
    \item \emph{Random: } \toolname performs better than the random baseline performs on all ten metrics for both datasets. 
    
    \item \emph{\toolname: } For coverage metrics, \toolname performs the third best for the Anime dataset and the second best for the YahooMusic dataset. It performs the best on five out of eight personalization metrics for both the datasets. (it is less than a two percentage points away from the best performing techniques for the remaining three personalization metrics for YahooMusic dataset). 
\end{enumerate}

\section{Evaluation Metrics for all architectures trained using the \underline{\textsc{MovieLens-100k}} dataset}
\label{sec:ml-100k-all-archs}

In this section, we experiment with all five architectures for the \underline{\textsc{MovieLens-100k}} dataset. The goal is to demonstrate the effectiveness of \toolname across a wide variety of collaborative filtering architectures. 
For each architecture, we train the best model using an extensive hyperparameter search. The metrics are reported in \Cref{tab:5-random-seeds-ml-100k-matrix_fact-repeat}, \Cref{tab:5-random-seeds-ml-100k-autoencoder}, \Cref{tab:5-random-seeds-ml-100k-ncf}, \Cref{tab:5-random-seeds-ml-100k-fm}, and \Cref{tab:5-random-seeds-ml-100k-dfm}. 
Across the five models, our conclusions regarding the comparison of \toolname and the baselines are similar to the ones presented in the evaluation section:
\begin{enumerate}
    \item \toolname performs better than LIME-RS across all five models and for all metrics
    \item \toolname performs better than AMCF-PH across all five models and for all metrics
    \item Global popularity performs better than \toolname for test set coverage and recommendation coverage, however \toolname performs better than it for all eight personalization metrics across all five models.
    \item User specific popularity performs second best for test set and recommendations coverage, however \toolname performs better than it for all eight personalization metrics across all five models. 
\end{enumerate}

\begin{table*}[h]
	\caption{The test set coverage, top-k recommendations coverage, and the 8 explanation personalization metrics are reported here (averaged over all users) for the \underline{\textsc{MovieLens-100k}} dataset trained using \underline{\textsc{Matrix Factorization}}. The mean and standard deviation value for each metric is computed over 5 random seeds. We compare 3 architecture choices for auxiliary modelling of RecXplainer. The best results are highlighted in bold. }
	\begin{center}
	\resizebox{1.0\textwidth}{!}{%
		\begin{tabular}{lllllllll}
			\toprule
			Metrics & LIME-RS & AMCF-PH & GBL Popl. & User Popl. & Random & RX-Linear & RX-MLP & RX-GBDT \\
			\midrule
			Testset Coverage & 57.42 $\pm$ 0.47 & 49.62 $\pm$ 4.59 & \textbf{84.69 $\pm$ 0.18} & 77.6 $\pm$ 0.59 & 32.83 $\pm$ 0.57 & 60.74 $\pm$ 0.21 & 67.16 $\pm$ 0.59 & 63.01 $\pm$ 0.3 \\
			Recommendations Coverage & 67.08 $\pm$ 2.68 & 44.2 $\pm$ 2.58 & \textbf{79.12 $\pm$ 2.83} & 70.12 $\pm$ 2.05 & 26.72 $\pm$ 0.89 & 69.3 $\pm$ 2.4 & 65.23 $\pm$ 2.28 & 64.27 $\pm$ 2.09 \\
			\midrule
			CondProb Generalpref Coverage & 59.85 $\pm$ 0.56 & 54.57 $\pm$ 1.73 & 25.07 $\pm$ 0.58 & 41.51 $\pm$ 0.44 & 45.07 $\pm$ 0.91 & 64.77 $\pm$ 1.75 & 60.78 $\pm$ 1.59 & \textbf{71.24 $\pm$ 1.12} \\
			CondProb Generalpref Ranking & 13.99 $\pm$ 0.41 & 13.74 $\pm$ 0.39 & 5.04 $\pm$ 0.16 & 9.76 $\pm$ 0.23 & 11.64 $\pm$ 0.29 & 15.8 $\pm$ 0.34 & 15.68 $\pm$ 0.17 & \textbf{20.28 $\pm$ 0.3} \\
			CondProb Specificpref Coverage & 48.87 $\pm$ 0.25 & \textbf{49.97 $\pm$ 1.14} & 25.1 $\pm$ 0.58 & 41.44 $\pm$ 0.44 & 43.94 $\pm$ 0.5 & 49.78 $\pm$ 0.48 & 49.06 $\pm$ 0.22 & \textbf{51.23 $\pm$ 0.27} \\
			CondProb Specificpref Ranking & 51.24 $\pm$ 0.15 & 49.89 $\pm$ 0.32 & 6.29 $\pm$ 0.11 & 10.29 $\pm$ 0.25 & 11.71 $\pm$ 0.14 & 51.62 $\pm$ 0.36 & 52.59 $\pm$ 0.27 & \textbf{55.98 $\pm$ 0.22} \\
			Odds Generalpref Coverage & 69.31 $\pm$ 1.43 & 60.98 $\pm$ 1.46 & 47.42 $\pm$ 0.55 & 55.78 $\pm$ 0.84 & 44.39 $\pm$ 0.49 & 74.42 $\pm$ 1.05 & 72.47 $\pm$ 1.16 & \textbf{81.15 $\pm$ 0.55} \\
			Odds Generalpref Ranking & 23.25 $\pm$ 0.58 & 18.94 $\pm$ 0.8 & 17.72 $\pm$ 0.16 & 27.63 $\pm$ 0.4 & 11.25 $\pm$ 0.32 & 25.87 $\pm$ 0.33 & 29.3 $\pm$ 0.83 & \textbf{33.57 $\pm$ 0.39} \\
			Odds Specificpref Coverage & 54.96 $\pm$ 0.62 & 51.8 $\pm$ 0.93 & 47.47 $\pm$ 0.55 & 55.73 $\pm$ 0.84 & 44.08 $\pm$ 0.24 & 56.17 $\pm$ 0.5 & 55.56 $\pm$ 0.6 & \textbf{57.74 $\pm$ 0.46} \\
			Odds Specificpref Ranking & 50.67 $\pm$ 0.16 & 50.07 $\pm$ 0.44 & 19.04 $\pm$ 0.32 & 28.04 $\pm$ 0.4 & 11.68 $\pm$ 0.1 & 50.34 $\pm$ 0.12 & 52.54 $\pm$ 0.28 & \textbf{55.33 $\pm$ 0.12} \\
			\bottomrule
		\end{tabular}
	}
	\end{center}
	\label{tab:5-random-seeds-ml-100k-matrix_fact-repeat}
\end{table*}

\begin{table*}[h]
	\caption{The test set coverage, top-k recommendations coverage, and the 8 explanation personalization metrics are reported here (averaged over all users) for the \underline{\textsc{MovieLens-100k}} dataset trained using \underline{\textsc{Autoencoder}}. The mean and standard deviation value for each metric is computed over 5 random seeds. We compare 3 architecture choices for auxiliary modelling of RecXplainer. The best results are highlighted in bold. }
	\begin{center}
	\resizebox{1.0\textwidth}{!}{%
		\begin{tabular}{lllllllll}
			\toprule
			Metrics & LIME-RS & AMCF-PH & GBL Popl. & User Popl. & Random & RX-Linear & RX-MLP & RX-GBDT \\
			\midrule
			Testset Coverage & 58.09 $\pm$ 0.55 & 47.91 $\pm$ 15.0 & \textbf{84.69 $\pm$ 0.18} & 77.88 $\pm$ 0.23 & 32.83 $\pm$ 0.57 & 60.52 $\pm$ 0.24 & 66.66 $\pm$ 1.16 & 64.22 $\pm$ 0.66 \\
			Recommendations Coverage & 64.06 $\pm$ 3.71 & 44.71 $\pm$ 21.94 & \textbf{79.07 $\pm$ 3.84} & 69.49 $\pm$ 2.87 & 25.24 $\pm$ 0.72 & 68.73 $\pm$ 3.95 & 65.58 $\pm$ 2.13 & 63.76 $\pm$ 2.68 \\
			\midrule
			CondProb Generalpref Coverage & 57.37 $\pm$ 1.76 & 44.52 $\pm$ 8.15 & 27.59 $\pm$ 0.69 & 42.42 $\pm$ 0.32 & 44.28 $\pm$ 1.26 & 65.51 $\pm$ 0.55 & 60.06 $\pm$ 0.42 & \textbf{70.56 $\pm$ 1.28} \\
			CondProb Generalpref Ranking & 13.47 $\pm$ 0.43 & 10.55 $\pm$ 3.35 & 5.55 $\pm$ 0.17 & 10.13 $\pm$ 0.2 & 11.21 $\pm$ 0.54 & 16.03 $\pm$ 0.35 & 14.67 $\pm$ 0.56 & \textbf{19.71 $\pm$ 0.7} \\
			CondProb Specificpref Coverage & 48.53 $\pm$ 0.37 & 44.84 $\pm$ 0.88 & 27.52 $\pm$ 0.69 & 42.36 $\pm$ 0.32 & 43.9 $\pm$ 0.28 & 49.96 $\pm$ 0.34 & 48.01 $\pm$ 0.29 & \textbf{50.98 $\pm$ 0.43} \\
			CondProb Specificpref Ranking & 50.85 $\pm$ 0.14 & 47.11 $\pm$ 1.77 & 6.8 $\pm$ 0.17 & 10.57 $\pm$ 0.19 & 11.65 $\pm$ 0.13 & 51.58 $\pm$ 0.3 & 51.6 $\pm$ 0.28 & \textbf{55.4 $\pm$ 0.27} \\
			Odds Generalpref Coverage & 67.95 $\pm$ 1.29 & 48.46 $\pm$ 4.32 & 47.38 $\pm$ 0.83 & 56.06 $\pm$ 0.98 & 44.05 $\pm$ 0.63 & 74.93 $\pm$ 0.67 & 70.97 $\pm$ 0.92 & \textbf{80.93 $\pm$ 1.83} \\
			Odds Generalpref Ranking & 22.25 $\pm$ 0.4 & 12.94 $\pm$ 2.23 & 18.2 $\pm$ 0.28 & 27.37 $\pm$ 0.48 & 11.2 $\pm$ 0.18 & 26.09 $\pm$ 0.41 & 27.33 $\pm$ 0.78 & \textbf{32.65 $\pm$ 0.61} \\
			Odds Specificpref Coverage & 55.08 $\pm$ 0.59 & 54.34 $\pm$ 0.65 & 47.43 $\pm$ 0.83 & 56.01 $\pm$ 0.98 & 44.01 $\pm$ 0.19 & 56.28 $\pm$ 0.55 & 54.84 $\pm$ 0.88 & \textbf{57.81 $\pm$ 0.63} \\
			Odds Specificpref Ranking & 50.39 $\pm$ 0.07 & 48.01 $\pm$ 1.13 & 19.76 $\pm$ 0.43 & 27.8 $\pm$ 0.5 & 11.66 $\pm$ 0.09 & 50.46 $\pm$ 0.17 & 51.89 $\pm$ 0.1 & \textbf{55.07 $\pm$ 0.14} \\
			\bottomrule
		\end{tabular}
	}
	\end{center}
	\label{tab:5-random-seeds-ml-100k-autoencoder}
\end{table*}

\begin{table*}[h]
	\caption{The test set coverage, top-k recommendations coverage, and the 8 explanation personalization metrics are reported here (averaged over all users) for the \underline{\textsc{MovieLens-100k}} dataset trained using \underline{\textsc{Neural CF}}. The mean and standard deviation value for each metric is computed over 5 random seeds. We compare 3 architecture choices for auxiliary modelling of RecXplainer. The best results are highlighted in bold. }
	\begin{center}
	\resizebox{1.0\textwidth}{!}{%
		\begin{tabular}{lllllllll}
			\toprule
			Metrics & LIME-RS & AMCF-PH & GBL Popl. & User Popl. & Random & RX-Linear & RX-MLP & RX-GBDT \\
			\midrule
			Testset Coverage & 23.26 $\pm$ 6.14 & 39.11 $\pm$ 10.17 & \textbf{84.69 $\pm$ 0.18} & 77.88 $\pm$ 0.23 & 32.83 $\pm$ 0.57 & 59.44 $\pm$ 0.92 & 66.6 $\pm$ 1.79 & 63.0 $\pm$ 0.46 \\
			Recommendations Coverage & 19.05 $\pm$ 4.81 & 44.14 $\pm$ 9.77 & \textbf{85.0 $\pm$ 0.0} & 76.31 $\pm$ 0.37 & 30.72 $\pm$ 0.41 & 64.4 $\pm$ 1.6 & 72.81 $\pm$ 2.36 & 67.25 $\pm$ 0.65 \\
			\midrule
			CondProb Generalpref Coverage & 43.9 $\pm$ 7.77 & 40.25 $\pm$ 4.08 & 27.59 $\pm$ 0.69 & 42.42 $\pm$ 0.32 & 44.28 $\pm$ 1.26 & \textbf{67.47 $\pm$ 0.99} & 53.36 $\pm$ 1.37 & \textbf{67.91 $\pm$ 1.21} \\
			CondProb Generalpref Ranking & 11.44 $\pm$ 3.02 & 9.63 $\pm$ 1.04 & 5.55 $\pm$ 0.17 & 10.13 $\pm$ 0.2 & 11.21 $\pm$ 0.54 & 16.63 $\pm$ 0.41 & 12.94 $\pm$ 0.39 & \textbf{18.25 $\pm$ 0.3} \\
			CondProb Specificpref Coverage & 44.08 $\pm$ 0.87 & 43.19 $\pm$ 1.77 & 27.52 $\pm$ 0.69 & 42.36 $\pm$ 0.32 & 43.9 $\pm$ 0.28 & \textbf{50.0 $\pm$ 0.4} & 46.6 $\pm$ 0.47 & \textbf{49.75 $\pm$ 0.35} \\
			CondProb Specificpref Ranking & 47.79 $\pm$ 0.46 & 47.52 $\pm$ 0.38 & 6.8 $\pm$ 0.17 & 10.57 $\pm$ 0.19 & 11.65 $\pm$ 0.13 & 51.83 $\pm$ 0.57 & 49.65 $\pm$ 0.32 & \textbf{53.47 $\pm$ 0.17} \\
			Odds Generalpref Coverage & 37.09 $\pm$ 2.15 & 42.95 $\pm$ 7.04 & 47.38 $\pm$ 0.83 & 56.06 $\pm$ 0.98 & 44.05 $\pm$ 0.63 & 74.89 $\pm$ 0.76 & 65.54 $\pm$ 1.78 & \textbf{77.45 $\pm$ 1.73} \\
			Odds Generalpref Ranking & 8.71 $\pm$ 0.71 & 11.3 $\pm$ 3.62 & 18.2 $\pm$ 0.28 & 27.37 $\pm$ 0.48 & 11.2 $\pm$ 0.18 & 25.62 $\pm$ 0.31 & 23.51 $\pm$ 1.07 & \textbf{29.33 $\pm$ 0.49} \\
			Odds Specificpref Coverage & 45.95 $\pm$ 1.0 & 48.66 $\pm$ 4.02 & 47.43 $\pm$ 0.83 & \textbf{56.01 $\pm$ 0.98} & 44.01 $\pm$ 0.19 & \textbf{55.03 $\pm$ 1.12} & 51.89 $\pm$ 1.04 & \textbf{55.48 $\pm$ 0.5} \\
			Odds Specificpref Ranking & 47.9 $\pm$ 0.2 & 48.19 $\pm$ 0.64 & 19.76 $\pm$ 0.43 & 27.8 $\pm$ 0.5 & 11.66 $\pm$ 0.09 & 50.45 $\pm$ 0.09 & 49.58 $\pm$ 0.18 & \textbf{52.64 $\pm$ 0.22} \\
			\bottomrule
		\end{tabular}
	}
	\end{center}
	\label{tab:5-random-seeds-ml-100k-ncf}
\end{table*}

\begin{table*}[h]
	\caption{The test set coverage, top-k recommendations coverage, and the 8 explanation personalization metrics are reported here (averaged over all users) for the \underline{\textsc{MovieLens-100k}} dataset trained using \underline{\textsc{Factorization Machine}}. The mean and standard deviation value for each metric is computed over 5 random seeds. We compare 3 architecture choices for auxiliary modelling of RecXplainer. The best results are highlighted in bold. }
	\begin{center}
	\resizebox{1.0\textwidth}{!}{%
		\begin{tabular}{lllllllll}
			\toprule
			Metrics & LIME-RS & AMCF-PH & GBL Popl. & User Popl. & Random & RX-Linear & RX-MLP & RX-GBDT \\
			\midrule
			Testset Coverage & 56.6 $\pm$ 1.04 & 75.69 $\pm$ 1.0 & \textbf{84.69 $\pm$ 0.18} & 77.88 $\pm$ 0.23 & 32.83 $\pm$ 0.57 & 60.87 $\pm$ 0.53 & 67.61 $\pm$ 1.8 & 63.51 $\pm$ 0.4 \\
			Recommendations Coverage & 62.8 $\pm$ 3.03 & \textbf{74.4 $\pm$ 2.39} & \textbf{79.12 $\pm$ 4.05} & 72.66 $\pm$ 2.11 & 31.88 $\pm$ 1.2 & 64.97 $\pm$ 3.29 & 68.41 $\pm$ 1.92 & 66.02 $\pm$ 2.78 \\
			\midrule
			CondProb Generalpref Coverage & 61.61 $\pm$ 1.14 & 38.58 $\pm$ 2.0 & 27.59 $\pm$ 0.69 & 42.42 $\pm$ 0.32 & 44.28 $\pm$ 1.26 & 64.5 $\pm$ 1.11 & 61.04 $\pm$ 1.36 & \textbf{71.37 $\pm$ 0.78} \\
			CondProb Generalpref Ranking & 15.68 $\pm$ 0.41 & 8.33 $\pm$ 0.42 & 5.55 $\pm$ 0.17 & 10.13 $\pm$ 0.2 & 11.21 $\pm$ 0.54 & 16.01 $\pm$ 0.39 & 15.21 $\pm$ 0.67 & \textbf{20.16 $\pm$ 0.47} \\
			CondProb Specificpref Coverage & 48.74 $\pm$ 0.3 & 47.55 $\pm$ 0.73 & 27.52 $\pm$ 0.69 & 42.36 $\pm$ 0.32 & 43.9 $\pm$ 0.28 & 49.95 $\pm$ 0.33 & 48.53 $\pm$ 0.24 & \textbf{51.15 $\pm$ 0.22} \\
			CondProb Specificpref Ranking & 51.41 $\pm$ 0.24 & 47.06 $\pm$ 0.48 & 6.8 $\pm$ 0.17 & 10.57 $\pm$ 0.19 & 11.65 $\pm$ 0.13 & 52.16 $\pm$ 0.29 & 51.75 $\pm$ 0.46 & \textbf{55.85 $\pm$ 0.13} \\
			Odds Generalpref Coverage & 71.54 $\pm$ 1.43 & 55.06 $\pm$ 1.42 & 47.38 $\pm$ 0.83 & 56.06 $\pm$ 0.98 & 44.05 $\pm$ 0.63 & 74.66 $\pm$ 1.19 & 71.9 $\pm$ 1.37 & \textbf{80.85 $\pm$ 0.77} \\
			Odds Generalpref Ranking & 23.73 $\pm$ 0.83 & 20.65 $\pm$ 0.55 & 18.2 $\pm$ 0.28 & 27.37 $\pm$ 0.48 & 11.2 $\pm$ 0.18 & 26.16 $\pm$ 0.33 & 28.17 $\pm$ 0.79 & \textbf{33.51 $\pm$ 0.51} \\
			Odds Specificpref Coverage & 55.15 $\pm$ 0.65 & 56.58 $\pm$ 0.55 & 47.43 $\pm$ 0.83 & 56.01 $\pm$ 0.98 & 44.01 $\pm$ 0.19 & 56.25 $\pm$ 0.52 & 55.09 $\pm$ 0.92 & \textbf{57.95 $\pm$ 0.63} \\
			Odds Specificpref Ranking & 50.77 $\pm$ 0.22 & 48.96 $\pm$ 0.28 & 19.76 $\pm$ 0.43 & 27.8 $\pm$ 0.5 & 11.66 $\pm$ 0.09 & 50.5 $\pm$ 0.13 & 52.08 $\pm$ 0.18 & \textbf{55.26 $\pm$ 0.09} \\
			\bottomrule
		\end{tabular}
	}
	\end{center}
	\label{tab:5-random-seeds-ml-100k-fm}
\end{table*}

\begin{table*}[h]
	\caption{The test set coverage, top-k recommendations coverage, and the 8 explanation personalization metrics are reported here (averaged over all users) for the \underline{\textsc{MovieLens-100k}} dataset trained using \underline{\textsc{Deep Factorization Machine}}. The mean and standard deviation value for each metric is computed over 5 random seeds. We compare 3 architecture choices for auxiliary modelling of RecXplainer. The best results are highlighted in bold. }
	\begin{center}
	\resizebox{1.0\textwidth}{!}{%
		\begin{tabular}{lllllllll}
			\toprule
			Metrics & LIME-RS & AMCF-PH & GBL Popl. & User Popl. & Random & RX-Linear & RX-MLP & RX-GBDT \\
			\midrule
			Testset Coverage & 54.82 $\pm$ 0.86 & 50.58 $\pm$ 0.77 & \textbf{84.69 $\pm$ 0.18} & 77.88 $\pm$ 0.23 & 32.83 $\pm$ 0.57 & 60.35 $\pm$ 2.11 & 68.37 $\pm$ 2.82 & 62.92 $\pm$ 0.53 \\
			Recommendations Coverage & 66.78 $\pm$ 1.17 & 51.93 $\pm$ 1.17 & \textbf{79.7 $\pm$ 2.13} & 73.2 $\pm$ 1.34 & 34.2 $\pm$ 0.7 & 67.51 $\pm$ 0.93 & 70.48 $\pm$ 1.57 & 65.93 $\pm$ 0.81 \\
			\midrule
			CondProb Generalpref Coverage & 53.76 $\pm$ 1.24 & 43.39 $\pm$ 2.34 & 27.59 $\pm$ 0.69 & 42.42 $\pm$ 0.32 & 44.28 $\pm$ 1.26 & 63.16 $\pm$ 1.19 & 53.72 $\pm$ 1.79 & \textbf{69.52 $\pm$ 1.25} \\
			CondProb Generalpref Ranking & 12.31 $\pm$ 0.55 & 9.33 $\pm$ 0.77 & 5.55 $\pm$ 0.17 & 10.13 $\pm$ 0.2 & 11.21 $\pm$ 0.54 & 15.1 $\pm$ 0.55 & 12.85 $\pm$ 0.68 & \textbf{19.11 $\pm$ 0.41} \\
			CondProb Specificpref Coverage & 47.21 $\pm$ 0.24 & 44.64 $\pm$ 0.89 & 27.52 $\pm$ 0.69 & 42.36 $\pm$ 0.32 & 43.9 $\pm$ 0.28 & \textbf{49.59 $\pm$ 0.68} & 47.13 $\pm$ 0.54 & \textbf{50.19 $\pm$ 0.26} \\
			CondProb Specificpref Ranking & 50.29 $\pm$ 0.11 & 48.69 $\pm$ 0.19 & 6.8 $\pm$ 0.17 & 10.57 $\pm$ 0.19 & 11.65 $\pm$ 0.13 & 51.69 $\pm$ 0.61 & 49.96 $\pm$ 0.26 & \textbf{54.08 $\pm$ 0.13} \\
			Odds Generalpref Coverage & 62.74 $\pm$ 1.56 & 52.81 $\pm$ 2.74 & 47.38 $\pm$ 0.83 & 56.06 $\pm$ 0.98 & 44.05 $\pm$ 0.63 & 73.23 $\pm$ 0.75 & 66.21 $\pm$ 1.2 & \textbf{80.11 $\pm$ 1.69} \\
			Odds Generalpref Ranking & 19.56 $\pm$ 0.39 & 15.96 $\pm$ 1.32 & 18.2 $\pm$ 0.28 & 27.37 $\pm$ 0.48 & 11.2 $\pm$ 0.18 & 25.37 $\pm$ 0.39 & 24.72 $\pm$ 0.36 & \textbf{30.97 $\pm$ 0.3} \\
			Odds Specificpref Coverage & 53.27 $\pm$ 0.62 & 50.5 $\pm$ 1.18 & 47.43 $\pm$ 0.83 & \textbf{56.01 $\pm$ 0.98} & 44.01 $\pm$ 0.19 & \textbf{55.47 $\pm$ 0.76} & 53.32 $\pm$ 0.74 & \textbf{56.42 $\pm$ 0.61} \\
			Odds Specificpref Ranking & 49.67 $\pm$ 0.12 & 49.36 $\pm$ 0.34 & 19.76 $\pm$ 0.43 & 27.8 $\pm$ 0.5 & 11.66 $\pm$ 0.09 & 50.37 $\pm$ 0.14 & 50.13 $\pm$ 0.19 & \textbf{53.79 $\pm$ 0.21} \\
			\bottomrule
		\end{tabular}
	}
	\end{center}
	\label{tab:5-random-seeds-ml-100k-dfm}
\end{table*}

\section{One feature removal vs. SHAP for feature attribution}
\label{sec:removal-based}

\begin{table*}[h]
	\caption{The test set coverage, top-k recommendations coverage, and the 8 explanation personalization metrics are reported here (averaged over all users) for the \underline{\textsc{MovieLens-100k}} dataset trained using \underline{\textsc{Matrix Factorization}} when the feature importances are computed using one feature removal and using SHAP. The mean and standard deviation value for each metric is computed over 5 random seeds. The best results are highlighted in bold. One feature removal achives similar results to SHAP, but is \textcolor{blue}{49} times faster.}
	\begin{center}
	\resizebox{1.0\textwidth}{!}{%
		\begin{tabular}{lllllll}
			\toprule
			Metrics & RX-Linear & RX-MLP & RX-GBDT & RX-Linear-SHAP & RX-MLP-SHAP & RX-GBDT-SHAP \\
			\midrule
			Testset Coverage & 60.74 $\pm$ 0.21 & \textbf{67.16 $\pm$ 0.59} & 63.01 $\pm$ 0.3 & 60.74 $\pm$ 0.21 & \textbf{67.26 $\pm$ 0.43} & 62.67 $\pm$ 0.53 \\
			Recommendations Coverage & \textbf{69.3 $\pm$ 2.4} & \textbf{65.23 $\pm$ 2.28} & 64.27 $\pm$ 2.09 & \textbf{69.3 $\pm$ 2.4} & \textbf{67.64 $\pm$ 2.28} & \textbf{65.73 $\pm$ 2.45} \\
			\midrule
			CondProb Generalpref Coverage & 64.77 $\pm$ 1.75 & 60.78 $\pm$ 1.59 & 71.24 $\pm$ 1.12 & 64.77 $\pm$ 1.75 & 62.76 $\pm$ 1.0 & \textbf{73.17 $\pm$ 0.39} \\
			CondProb Generalpref Ranking & 15.8 $\pm$ 0.34 & 15.68 $\pm$ 0.17 & 20.28 $\pm$ 0.3 & 15.8 $\pm$ 0.34 & 16.47 $\pm$ 0.28 & \textbf{20.95 $\pm$ 0.34} \\
			CondProb Specificpref Coverage & 49.78 $\pm$ 0.48 & 49.06 $\pm$ 0.22 & \textbf{51.23 $\pm$ 0.27} & 49.78 $\pm$ 0.48 & 50.43 $\pm$ 0.16 & \textbf{51.83 $\pm$ 0.36} \\
			CondProb Specificpref Ranking & 51.62 $\pm$ 0.36 & 52.59 $\pm$ 0.27 & 55.98 $\pm$ 0.22 & 51.62 $\pm$ 0.36 & 53.67 $\pm$ 0.37 & \textbf{56.69 $\pm$ 0.3} \\
			Odds Generalpref Coverage & 74.42 $\pm$ 1.05 & 72.47 $\pm$ 1.16 & 81.15 $\pm$ 0.55 & 74.42 $\pm$ 1.05 & 74.02 $\pm$ 0.93 & \textbf{83.18 $\pm$ 0.27} \\
			Odds Generalpref Ranking & 25.87 $\pm$ 0.33 & 29.3 $\pm$ 0.83 & \textbf{33.57 $\pm$ 0.39} & 25.87 $\pm$ 0.33 & 30.8 $\pm$ 0.84 & \textbf{34.2 $\pm$ 0.34} \\
			Odds Specificpref Coverage & 56.17 $\pm$ 0.5 & 55.56 $\pm$ 0.6 & \textbf{57.74 $\pm$ 0.46} & 56.17 $\pm$ 0.5 & \textbf{57.8 $\pm$ 0.55} & \textbf{58.41 $\pm$ 0.46} \\
			Odds Specificpref Ranking & 50.34 $\pm$ 0.12 & 52.54 $\pm$ 0.28 & 55.33 $\pm$ 0.12 & 50.34 $\pm$ 0.12 & 53.14 $\pm$ 0.28 & \textbf{55.69 $\pm$ 0.12} \\
			\bottomrule
		\end{tabular}
	}
	\end{center}
	\label{tab:5-random-seeds-ml-100k-matrix_fact-one-feature-removal-vs-shap}
\end{table*}

\begin{table*}[h]
	\caption{The test set coverage, top-k recommendations coverage, and the 8 explanation personalization metrics are reported here (averaged over all users) for the \underline{\textsc{Hetrec}} dataset trained using \underline{\textsc{Matrix Factorization}} when the feature importances are computed using one feature removal and using SHAP. The mean and standard deviation value for each metric is computed over 5 random seeds. The best results are highlighted in bold. One feature removal achives similar results to SHAP, but is \textcolor{blue}{90} times faster.}
	\begin{center}
	\resizebox{1.0\textwidth}{!}{%
		\begin{tabular}{lllllll}
			\toprule
			Metrics & RX-Linear & RX-MLP & RX-GBDT & RX-Linear-SHAP & RX-MLP-SHAP & RX-GBDT-SHAP \\
			\midrule
			Testset Coverage & 75.46 $\pm$ 0.34 & \textbf{80.98 $\pm$ 1.18} & 78.77 $\pm$ 0.18 & 75.45 $\pm$ 0.33 & \textbf{82.13 $\pm$ 1.43} & 78.16 $\pm$ 0.24 \\
			Recommendations Coverage & \textbf{81.4 $\pm$ 1.22} & \textbf{79.65 $\pm$ 2.59} & 78.98 $\pm$ 0.88 & \textbf{81.4 $\pm$ 1.22} & \textbf{82.55 $\pm$ 2.38} & \textbf{81.25 $\pm$ 0.95} \\
			\midrule
			CondProb Generalpref Coverage & 84.61 $\pm$ 0.42 & 78.63 $\pm$ 2.16 & 84.56 $\pm$ 0.45 & 84.63 $\pm$ 0.44 & 76.71 $\pm$ 1.3 & \textbf{85.75 $\pm$ 0.63} \\
			CondProb Generalpref Ranking & 17.72 $\pm$ 0.19 & 16.4 $\pm$ 0.62 & \textbf{19.55 $\pm$ 0.37} & 17.73 $\pm$ 0.18 & 15.98 $\pm$ 0.63 & \textbf{20.06 $\pm$ 0.26} \\
			CondProb Specificpref Coverage & \textbf{63.52 $\pm$ 0.45} & 59.87 $\pm$ 0.45 & 61.86 $\pm$ 0.47 & \textbf{63.52 $\pm$ 0.45} & 60.19 $\pm$ 0.83 & \textbf{63.06 $\pm$ 0.46} \\
			CondProb Specificpref Ranking & 54.26 $\pm$ 0.06 & 54.13 $\pm$ 0.47 & 57.46 $\pm$ 0.09 & 54.26 $\pm$ 0.06 & 55.76 $\pm$ 0.5 & \textbf{58.69 $\pm$ 0.1} \\
			Odds Generalpref Coverage & \textbf{85.15 $\pm$ 0.5} & 79.01 $\pm$ 2.28 & 84.79 $\pm$ 0.53 & \textbf{85.16 $\pm$ 0.49} & 76.7 $\pm$ 0.81 & \textbf{85.85 $\pm$ 0.49} \\
			Odds Generalpref Ranking & 23.32 $\pm$ 0.28 & 23.08 $\pm$ 0.48 & 26.0 $\pm$ 0.22 & 23.32 $\pm$ 0.28 & 23.64 $\pm$ 0.36 & \textbf{26.72 $\pm$ 0.3} \\
			Odds Specificpref Coverage & \textbf{65.78 $\pm$ 0.42} & 63.05 $\pm$ 0.38 & 64.86 $\pm$ 0.29 & \textbf{65.78 $\pm$ 0.42} & 63.79 $\pm$ 0.13 & \textbf{65.83 $\pm$ 0.32} \\
			Odds Specificpref Ranking & 53.49 $\pm$ 0.14 & 53.36 $\pm$ 0.35 & 56.35 $\pm$ 0.04 & 53.49 $\pm$ 0.14 & 54.91 $\pm$ 0.29 & \textbf{57.68 $\pm$ 0.08} \\
			\bottomrule
		\end{tabular}
	}
	\end{center}
	\label{tab:5-random-seeds-hetrec-matrix_fact-one-feature-removal-vs-shap}
\end{table*}

\citet{Ians-paper-xai-by-removal-survey} developed a framework to categorize such methods along three dimensions: 
\begin{itemize}[leftmargin=*,itemsep=2pt]
    \item Attribute removal: how the approach removes attributes from the model, 
    \item Model behavior: what model behavior is it observing, and 
    \item Summary technique: how does it summarize an attribute's impact? 
\end{itemize}

\toolname, when instantiated in this framework:
removes attributes by setting them to zero, 
analyzes prediction as the model behavior, and 
summarizes an attribute's impact by removing them individually. 

Since our attribute input vector is binary and indicates whether an attribute is present or not, simulating an attribute's removal by setting it to zero is a natural choice. 
Previous removal-based explanation methods have used \emph{prediction} or \emph{prediction loss} or \emph{dataset loss} for model behavior analysis. We chose \emph{prediction} instead of \emph{prediction loss} for our analysis because we wanted to get specific preference (analog of local interpretability) without requiring the original rating. 
Previous removal-based explanation methods have used \emph{removing individual attributes} or \emph{Shapley values} or \emph{trained additive models} to get attribute impact value. Removing individual attributes accesses the impact of an attribute by measuring the loss in prediction when that one attribute is removed, and this is what we choose. 
On the other hand, Shapley value takes all subsets of attributes and then use the cooperative game theoretic formulation to assign impact value to each attribute. It has two disadvantages: 
\begin{itemize}[leftmargin=*]
    \item It creates all subsets of attributes -- which is exponential in the number of attributes, making the process very expensive. 
    \item For creating all the subsets, it simulates removing many features that can potentially create attribute vectors that the auxiliary model has not seen, and thereby its prediction can not be trusted in that part of the data manifold. 
\end{itemize}
\toolname is by design agnostic to the specific feature removal technique. 
\Cref{tab:5-random-seeds-ml-100k-matrix_fact-one-feature-removal-vs-shap} and \Cref{tab:5-random-seeds-hetrec-matrix_fact-one-feature-removal-vs-shap} show the comparison of using one feature removal and SHAP for feature attribution for \underline{\textsc{MovieLens-100k}} and \underline{\textsc{Hetrec}} datasets trained using \underline{\textsc{Matrix Factorization}} models. 
In both the cases, we find that SHAP attains slightly better values for most metrics, however, it is much more expensive than using one feature removal technique (\textbf{49X} and \textbf{90X} respectively). 
For this reason, we choose the method of removing one feature as the default choice for feature attribution in \toolname. 


\section{Ablation Experiments}
\label{sec:ablation-expt-plots}

\begin{figure*}
    \centering
    \includegraphics[width=0.45\textwidth]{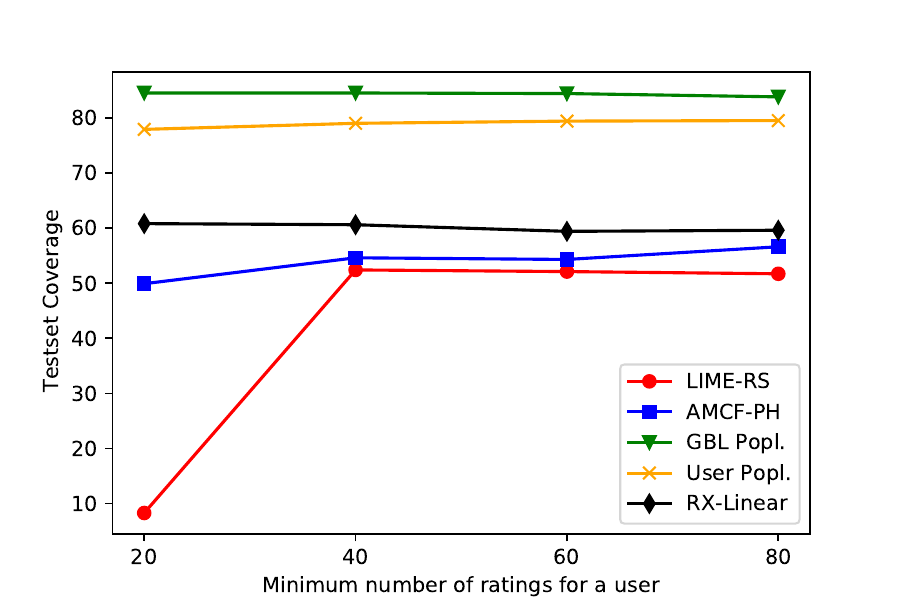}
    \includegraphics[width=0.45\textwidth]{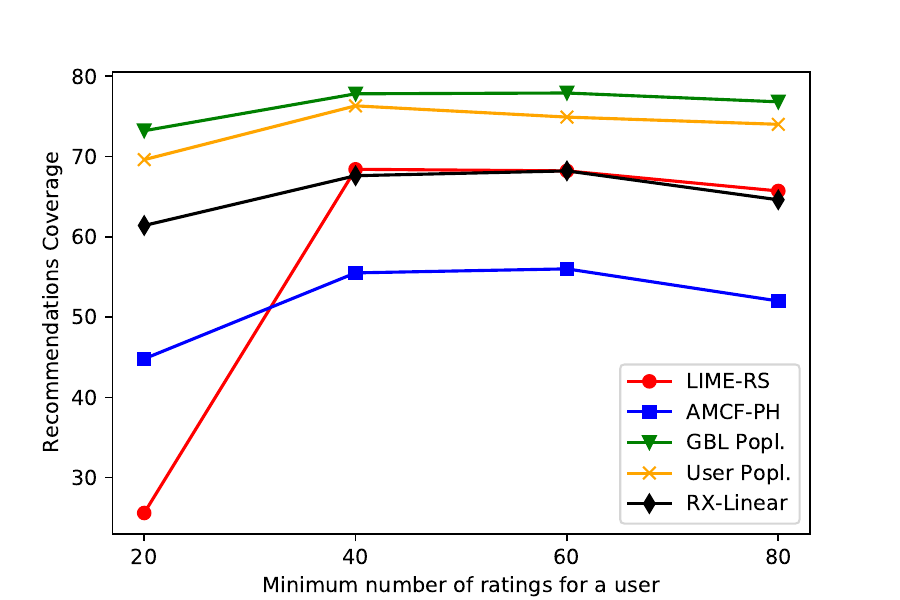}
    \includegraphics[width=0.45\textwidth]{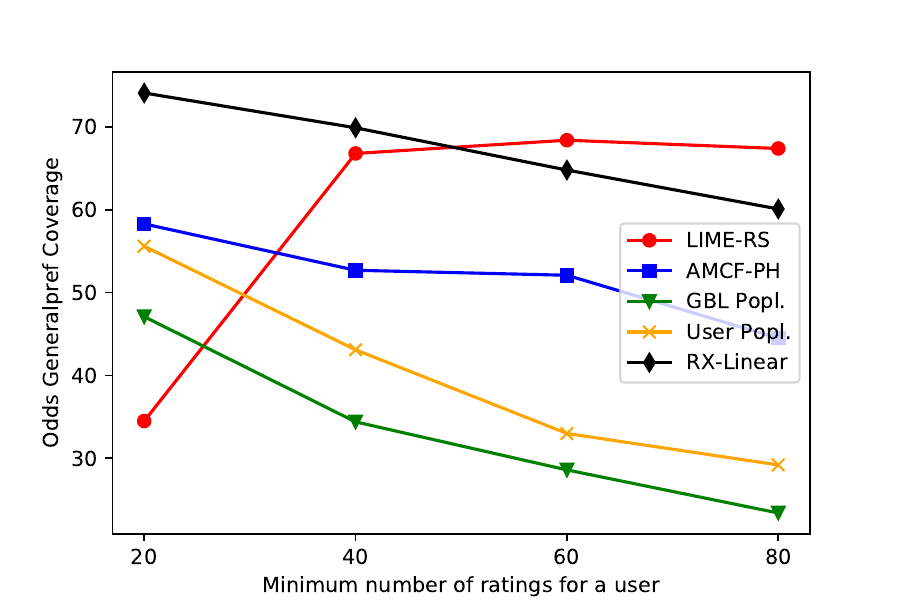}
    \includegraphics[width=0.45\textwidth]{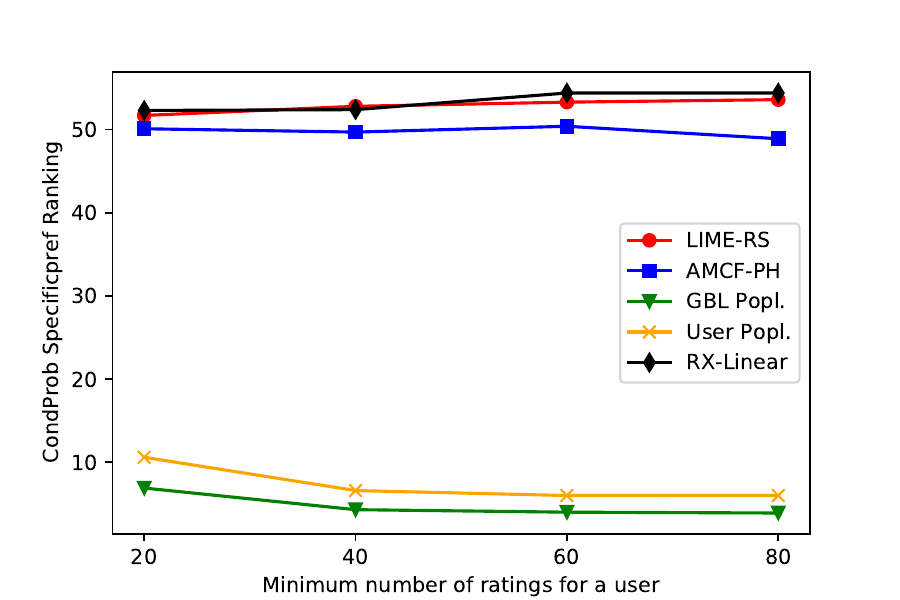}
    \caption{Ablation experiments on Movielens 100k dataset. We only retain users that have rated a minimum of 20, 40, 60, or 80 ratings. The first row shows the test set coverage and recommendations coverage. \toolname performs consistently ranks the third best across all the minimum user ratings. The second row shows the general preference coverage and ranking of specific preferences. \toolname ranks first for specific preference ranking across all minimum user ratings. For general preference coverage, \toolname ranks the best at lower minimum user ratings and second best afterwards. }
    \label{fig:ablation_increasing_min_user_rating_ml-100k}
\end{figure*}

\begin{figure*}
    \centering
    \includegraphics[width=0.45\textwidth]{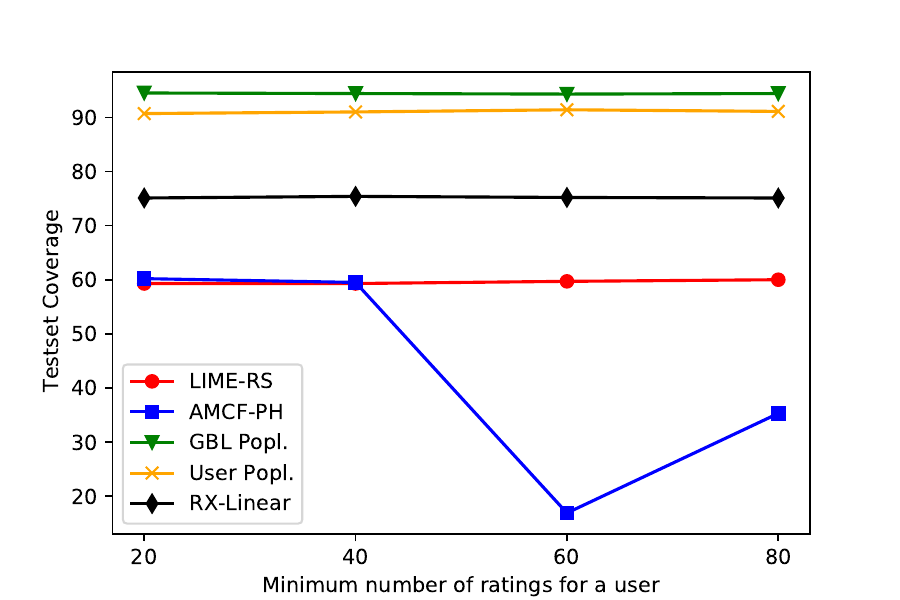}
    \includegraphics[width=0.45\textwidth]{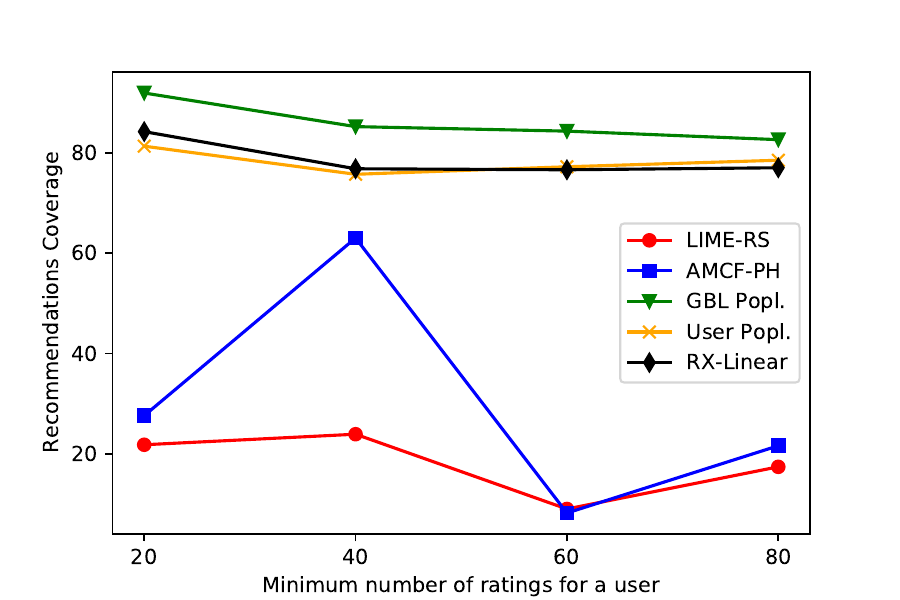}
    \includegraphics[width=0.45\textwidth]{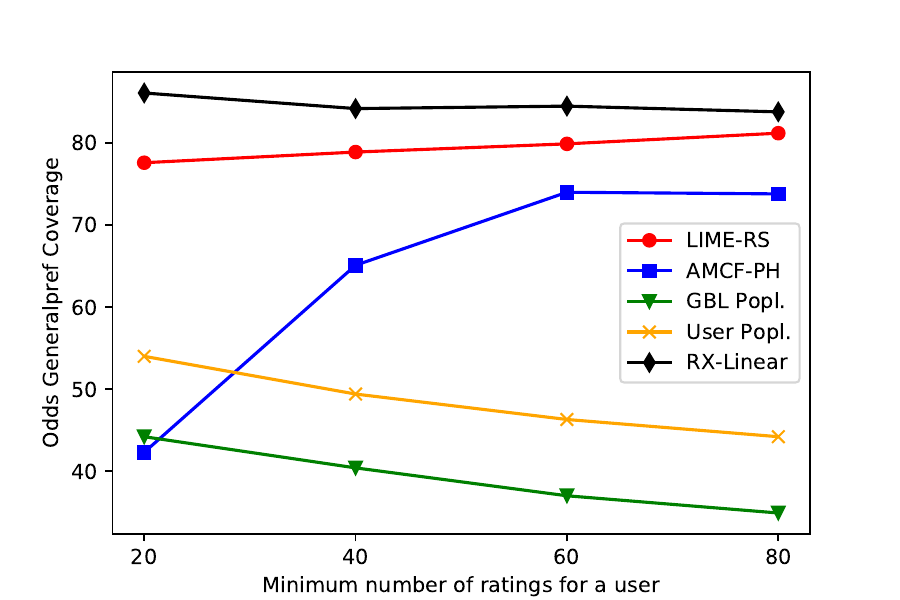}
    \includegraphics[width=0.45\textwidth]{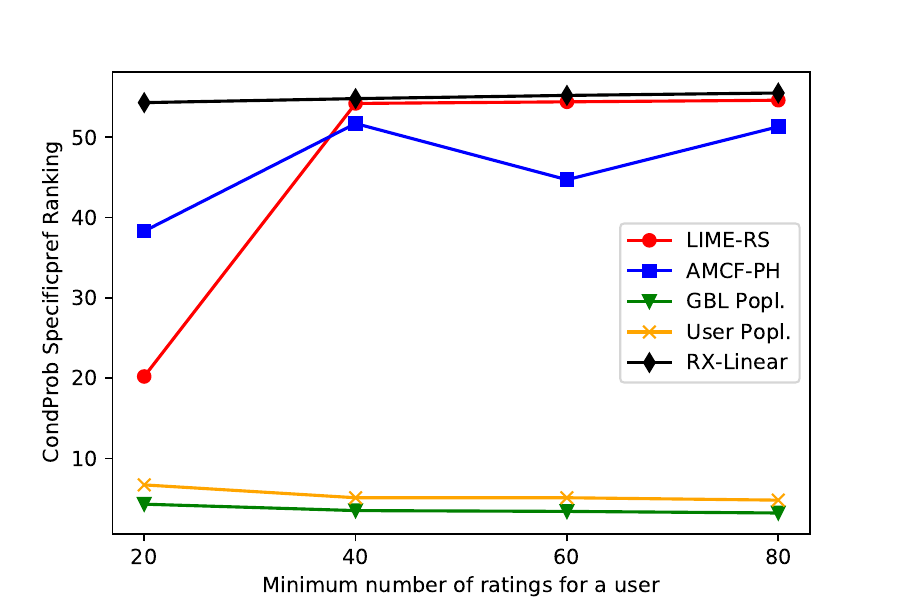}
    \caption{Ablation experiments on Hetrec Movielens dataset. We only retain users that have rated a minimum of 20, 40, 60, or 80 ratings. The first row shows the test set coverage and recommendations coverage. \toolname performs consistently ranks the third best across all the minimum user ratings for test set coverage. For recommendations coverage, it is the second best for settings 20 and 40, and the third best for settings 60 and 80. The second row shows the general preference coverage and ranking of specific preferences. \toolname ranks first consistently across all settings. }
    \label{fig:ablation_increasing_min_user_rating_hetrec}
\end{figure*}

We also performed ablation experiments by removing users who rated few items. More ratings per user means that the estimation of proxy measures will be more accurate as attributes would have a higher probability of prevalence in the liked and rated item sets. Thereby, we reduce the uncertainty in the proxy measures by only considering users that rate more items than a threshold. 
We constructed different versions of the \underline{\textsc{Movielens-$100K$}} and \underline{\textsc{Hetrec}} datasets where we filter users to have a minimum rating of different thresholds: 20, 40, 60, or 80. We then reran the experiments for these versions of the dataset and computed all the metrics. We report the trends in these metrics in \Cref{fig:ablation_increasing_min_user_rating_ml-100k} for \underline{\textsc{Movielens-$100K$}} and \Cref{fig:ablation_increasing_min_user_rating_hetrec} for \underline{\textsc{Hetrec}} dataset respectively. For \toolname, we show the metrics using the linear layer as its auxiliary model. 

\textbf{\underline{\textsc{Movielens-$100K$}}} dataset:
For all the minimum rating settings, \toolname performs the third best for the test-set and top-$k$ recommendations coverage-based metrics (behind global and user-specific popularity). For specific preference ranking, it performs the best across all settings. It performs the best for general preference coverage at lower minimum user ratings (20 and 40) and second for higher minimum user ratings (60 and 80). 

\textbf{\underline{\textsc{Hetrec}}} dataset:
For all the minimum rating settings, \toolname performs the third best for test set coverage. For recommendations coverage, it performs second best for minimum rating settings 20 and 40 and third best for minimum rating settings 60 and 80. 
For specific preference ranking and general preference coverage, \toolname consistently performs the best across all settings. 

Therefore, we can conclude that \toolname performs reliably well across a wide set of minimum user rating settings. 

%

\section{AMCF Evaluation Metrics:}
\label{sec:amcf-metrics-bad}
AMCF also evaluate their general and specific preference. For general preference, similar to our evaluation metrics, their primary metric is \emph{Top M recall at K}, i.e., the intersection between the top M ground truth user preference and top K predicted user preferences. 
To realize this metric, they need the ground truth user preference -- which is not present in the dataset -- and hence similar to our technique, they need to simulate. The way they compute the ground truth preference is unjustified and inexplicable. 
The process involves computing the weight of each item by removing the user and item bias terms, and then the preference of an attribute is just the sum of the weights of the items it occurs in. The latter part is still reasonable; however, we are uncertain about the weight calculation part. 
For evaluating specific preferences, they use the sorted order of general preference for the attributes present in that specific item -- which does not resolve the concerns mentioned above. 

We also use proxies to simulate the users' ground truth attribute preference; however there are key differences in our evaluation when compared to AMCF:
\begin{itemize}[leftmargin=*,itemsep=2pt]
    \item The proxies are more generalizable and reasonable, like the conditional probability of liking and odds of liking. 
    \item We use multiple proxies and report results on all of them to avoid cherry-picking. 
\end{itemize}

\section{AMCF Post-Hoc Adaptation:}
\label{sec:amcf-post-hoc-adapt}
\begin{figure}[h]
    \centering
    \includegraphics[width=\columnwidth]{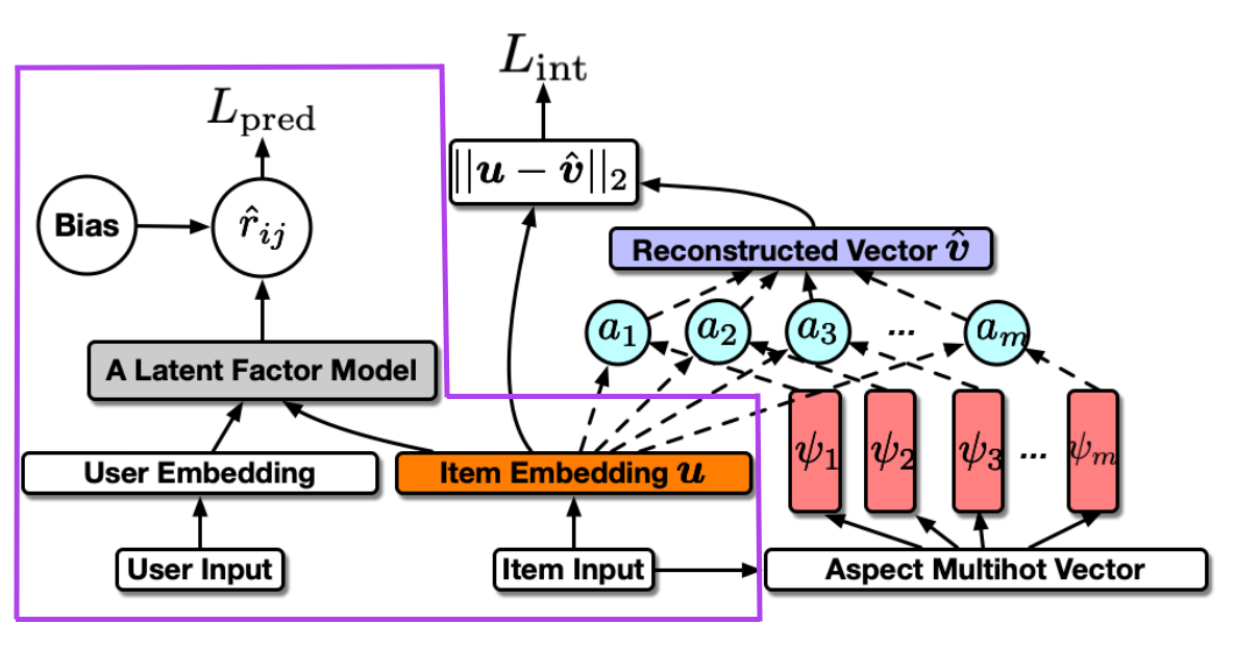}
    \caption{The architecture of AMCF Post-Hoc (AMCF-PH). The part within the blue colored region was not trained, thereby making the technique post-hoc. }
    \label{fig:AMCF-PH}
\end{figure}

For adapting AMCF~\citep{amcf-paper} to be post-hoc, we made a minor modification to the architecture mentioned in the paper (see \Cref{fig:AMCF-PH}). We froze the left-hand side of the architecture and only trained the right-hand side, which consists of the attention network that aims to reconstruct the item's embedding from its attributes. Since the user and item embeddings were not trained, this made the technique post-hoc. 

\section{Justification of using Rank-Biased Overlap (RBO):}
\label{sec:rbo-just}
Comparing ranked lists can be achieved using various rank correlation metrics like Kendall's Tau and Spearman's correlation~\citep{rank-correlation}. These metrics have restrictions that the ranked lists must be conjoint, i.e., they both must contain all the items that the entire universe of items contains. This is not suitable for our metrics because the top-$k$ attributes from either the conditional probability or the odds liking proxy might not have any common element with the ranking produced by any technique. Hence we choose to compare the ranked lists using rank-biased overlap (RBO) metric~\citep{RBO-paper}, which was recently proposed to overcome the limitation posed by correlation-based metrics. Specifically, RBO is better than correlation-based metrics as:
\begin{itemize}
    \item RBO does not require the two ranked lists to be conjoint, i.e., a ranked list having items that do not occur in the other list is acceptable, e.g., the similarity between [1, 3, 7] and [1, 5, 8] can be measured using RBO. 
    \item RBO does not require the two ranked lists to be of the same length. 
    \item RBO provides weighted comparison, i.e., discordance at higher ranks is penalized more than discordance at lower ranks. 
\end{itemize}

\end{document}